\documentclass[useAMS,usenatbib]{mn2e}
\usepackage[]{graphicx}
\usepackage{psfig}
\usepackage{rotating}
\usepackage{epsfig}
\usepackage{fixltx2e}
\usepackage{appendix}
\usepackage{psfrag}
\usepackage{amsmath,amssymb}
\usepackage{threeparttable}
\usepackage{float}
\usepackage{longtable}
\usepackage{lscape}
\usepackage{afterpage}
\usepackage{url}
\usepackage{color}

\def \oiii {[O{\small~III}]}
\def \oii {[O{\small~II}]}
\def \oi {[O{\small~I}]}
\def \sii {[S{\small~II}]}
\def \nii {[N{\small~II}]}
\def \ha {H$\alpha$}
\def \hb {H$\beta$}
\def \hi {H{\sc i}}
\def \Msol {$M_{\odot}$}

\title[]{The SAMI Galaxy Survey: instrument specification and target selection}

\author[J. J. Bryant et al.]{J. J. Bryant$^{1,2,3}$\thanks{E-mail: jbryant@physics.usyd.edu.au (JJB)}, M. S. Owers$^{2}$, A. S. G. Robotham$^{4}$, S. M. Croom$^{1,3}$, 
\newauthor %
S. P. Driver$^{4,6}$, M. J. Drinkwater$^{5}$, N. P. F. Lorente$^{2}$, L. Cortese$^{8}$, N. Scott$^{1,3}$,  \newauthor
M. Colless$^{9}$, A. Schaefer$^{1,2,3}$, E. N. Taylor$^{11}$, I. S. Konstantopoulos$^{2,3}$,  \newauthor 
J. T. Allen$^{1,3}$, I. Baldry$^{7}$, L. Barnes$^{1}$, A. E. Bauer$^{2}$, J. Bland-Hawthorn$^{1,3,13}$, \newauthor 
J. V. Bloom$^{1,3}$, A. M. Brooks$^{15}$,  S. Brough$^{2}$, G. Cecil$^{18,22}$, W. Couch$^{2}$,   D. Croton$^{8}$, \newauthor R. Davies$^{16}$, S. Ellis$^{2}$, L. M. R. Fogarty$^{1,3}$,  C. Foster$^{2}$, K. Glazebrook$^{8}$, \newauthor M. Goodwin$^{2}$,  A. Green$^{2}$, M. L. Gunawardhana$^{17}$, E. Hampton$^{9}$, I.-T. Ho$^{12}$, \newauthor A. M. Hopkins$^{2}$,  L. Kewley$^{9}$, J. S. Lawrence$^{2}$,  S. G. Leon-Saval$^{13}$, S. Leslie$^{9}$, \newauthor R. McElroy$^{1,3}$, G. Lewis$^{1}$,  J. Liske$^{10}$,   \'A.R. L\'opez-S\'anchez$^{2,20}$, S. Mahajan$^{19,5}$, \newauthor A. M. Medling$^{9}$,  N. Metcalfe$^{21}$,  M. Meyer$^{4}$, J. Mould$^{8}$,  D. Obreschkow$^{4}$, \newauthor S. O'Toole$^{2}$,  M. Pracy$^{1}$, S. N. Richards $^{1,2,3}$, T. Shanks$^{21}$,  R. Sharp$^{9,3}$,  \newauthor S. M. Sweet$^{5,9}$,   A. D. Thomas$^{5}$,  C. Tonini$^{11}$, C. J. Walcher$^{14}$
 \\
{\it Affiliations can be found after the references}\\}

\begin{document}
\date{}
\pagerange{\pageref{firstpage}--\pageref{lastpage}} \pubyear{2011}
\maketitle

\label{firstpage}

\begin{abstract}

The SAMI Galaxy Survey will observe 3400 galaxies with the Sydney-AAO Multi-object Integral-field spectrograph (SAMI) on the Anglo-Australian Telescope (AAT) in a 3-year survey which began in 2013. We present the throughput of the SAMI system, the science basis and specifications for the target selection, the survey observation plan and the combined properties of the selected galaxies. The survey includes four volume-limited galaxy samples based on cuts in a proxy for stellar mass, along with low-stellar-mass dwarf galaxies all selected from the Galaxy And Mass Assembly (GAMA) survey. 
The GAMA regions were selected because of the vast array of ancillary data available, including ultraviolet through to radio bands.
These fields are on the celestial equator at 9, 12, and 14.5 hours, and cover a total of 144 square degrees (in GAMA-I).  Higher density environments are also included with the addition of eight clusters. The clusters have spectroscopy from 2dFGRS and SDSS and photometry in regions covered by the Sloan Digital Sky Survey (SDSS) and/or VLT Survey Telescope/ATLAS. The aim is to cover a broad range in stellar mass and environment, and therefore the primary survey targets cover redshifts $0.004 < z < 0.095$, magnitudes $r_{pet} < 19.4$, stellar masses $10^{7} $-- $10^{12}$\Msol, and environments from isolated field galaxies through groups to clusters of $\sim10^{15}$\Msol. 

\end{abstract}


\section{Introduction}
\label{intro}

Over the last two decades, significant advances in our understanding of galaxies have been driven by large galaxy surveys such as the Sloan Digital Sky Survey \citep[SDSS;][]{Yor2000, Aba09}, 2-degree Field Galaxy Redshift Survey  \citep[2dFGRS;][]{Col01}, the Cosmic Evolution Survey \citep[COSMOS;][]{Sco07}, the VIMOS VLT Deep Survey \citep[VVDS;][]{LeF04} and the Galaxy and Mass Assembly (GAMA) survey \citep{Dri09, Dri11}. While these surveys have resulted in more than a 3.5 million galaxy spectra, most of these spectra have been taken with a single spectrum (fibre or slit). The spectrum is susceptible to aperture effects as it records a different fraction and part of the galaxy depending on the size or distance of the galaxy and the positioning of the fibre. Integral-field unit (IFU) spectroscopy, on the other hand, spatially resolves each galaxy, giving a spectrum at multiple positions across the galaxy. The gain in information from IFUs over that of single-fibre surveys include the spatial distribution of gas and star formation, kinematic information revealing the mass (and dark matter) distributions as well as tracing regularity or disturbance in gas or stellar motions, gradients across the galaxy in stellar and/or gas metallicity and age, and resolved emission lines to map the processes driving ionisation in different parts of the galaxy. Disentangling the relationships between all of these observables and the galaxy mass, redshift (evolution) and environment, requires a large sample size. Such a large sample can only be achieved with a multiplex IFU instrument in which many IFUs in the focal plane greatly increase the speed of a galaxy survey.

Several IFU galaxy surveys have begun in the last few years. While a number of IFU surveys have covered tens of objects, such as VENGA \citep{Bla13}, VIXENS \citep{Hei11} and PINGS \citep{Ros10}, many hundreds of galaxies are required to divide the parameter space. The first survey to amass a significant sample of galaxies was the SAURON survey \citep{deZ02} of 72 nearby galaxies, which was then extended into the ATLAS$^{\rm 3D}$ survey \citep{Cap11} and resulted in $260$ galaxies at $z < 0.01$. These galaxies were observed with the SAURON integral-field spectrograph on the 4.2m William Herschel Telescope using a resolution of $R \sim 1200$ and a field-of-view of $33 \times 41$\,arcsec.  Until recently, the largest IFU survey underway was the CALIFA survey \citep{San12, Wal14} using the Potsdam Multi-Aperture Spectrophotometer \citep[PMAS;][]{Rot05} IFU on the 3.5m Calar Alto Telescope. CALIFA comprises a total of 600 galaxies to $z < 0.03$ at resolutions of $R\sim 850$ and $1650$ in the blue and red respectively, and a field of view of $74 \times 64$ arcsec. While both the ATLAS$^{\rm 3D}$ and CALIFA surveys have large fields of view, the instruments  do not have the multiplexing required to easily reach thousands of galaxies.

The Sydney-AAO Multi-object Integral-field spectrograph \citep[SAMI;][]{Cro2012, JB2012} achieves this multiplexing using revolutionary new imaging fibre bundles, called {\it hexabundles} \citep{JB2014, JB2011, JBH2011}. Each hexabundle has 61 optical fibres with cores that subtend 1.6 arcsec on the sky, giving a total hexabundle diameter of 15 arcsec, and physical size $< 1$mm, with a filling fraction of 73\%. Thirteen of these hexabundles manually plug into a field plate with pre-drilled holes, which is installed at the prime focus of the Anglo-Australian Telescope (AAT). This instrument allows simultaneous IFU observations of 12 galaxies and one calibration star, significantly increasing the rate at which galaxy observations can be collected compared with single IFU instruments. 

The SAMI instrument began taking pilot data for the SAMI galaxy survey in 2012, continuing on to the main galaxy survey in 2013. The fundamental aim is a survey of 3400 galaxies across a broad range of environments and stellar masses.  
The SAMI Galaxy Survey will observe an order of magnitude more galaxies than any previous IFU surveys.
 SAMI feeds the AAOmega spectrograph  \citep{Sha06}, which for the survey is set up to have resolutions of $R = 1730$ in the blue arm and $R = 4500$ in the red arm. The unique challenges of the SAMI data reduction are discussed in \citet{Sha14}, and the details of the SAMI early data release are given in \citet{All14}.
 
Where we refer to SAMI galaxies or the SAMI sample we are referring to the galaxies or sample of the SAMI Galaxy Survey.

The outline of this paper is as follows: Section~\ref{instrument} discusses the improvements in the SAMI instrument as a result of the upgrade from SAMI-I to SAMI-II; Sections~\ref{TS} and~\ref{constraints} present the main science drivers influencing the target selection and the constraints on the target selection to meet those science drivers respectively; the final definition of the selected sample for the field and group galaxies from GAMA is given in Section~\ref{GAMA}, and for the cluster galaxies, is in Section~\ref{Clusters}; Section~\ref{sec_comb} illustrates the combined sample properties; Section~\ref{ancil} discusses ancillary data available at other wavelengths; and finally Section~\ref{Observing} outlines how the SAMI galaxy survey proceeds including guide and standard star selection, and the {\it Greedy} tiling algorithm.
Throughout this paper, we adopt the concordance cosmology:
($\Omega_{\Lambda}$,$\Omega_{\rm m}$, h) $=$ (0.7, 0.3, 0.7) \citep{Hin09}.
Colour versions of all figures appear in the online version.

\section{The upgraded SAMI instrument}
\label{instrument}

Ahead of the start of the SAMI Galaxy Survey in semester 2013A, the original SAMI instrument as detailed in \citet{Cro2012}, was upgraded. The improvements included:
new design for the connectors to attach the hexabundles to the field plate;
new and improved hexabundles fabricated at the University of Sydney;
change in the fibre type used, in order to increase the blue throughput;
new cabling for the 42m of optical fibre that runs from the telescope top end, down to the coud\'{e} room, to reduce focal ratio degradation (FRD).

\subsection{New hexabundle connectors}

\begin{figure}
\centerline{\psfig{file=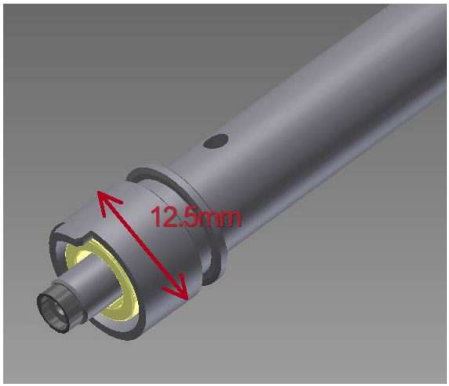, width=8.0cm}}
\vspace*{0.5mm}
\centerline{\psfig{file=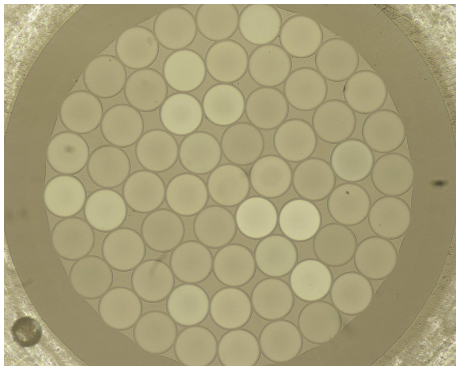, width=6.0cm}}
\vspace*{3mm}
\caption{Top: New design for the magnetic connector which attaches a hexabundle to the field plate. The diameter of the footprint of the connector is 12.5mm on the field plate, requiring a separation of at least 15mm between galaxies in any one tiled field. A rectangular protrusion or `key' in the outer ring slots into a smaller hole in the plate beside a larger hole for the central hexabundle ferule. The key secures the rotation of the bundle relative to the plate. Lower: One of the 61-core hexabundles manufactured at the University of Sydney. The diameter of the hexabundle is $< 1$mm and is mounted in the centre of the smallest tube in the top image, inset in the black cylinder which protects the face of the fibres.  
}
\label{connectors}
\end{figure}

The SAMI-I prototype instrument had the hexabundles mounted in off-the-shelf connectors that attached to the field plate in the top end of the AAT. The main issue with this design was that the orientation of the connectors was not fixed to any reference, leading to galaxy images that were randomly oriented on the sky. New connectors were designed and fitted with our new hexabundles, and are shown in Figure~\ref{connectors}. The hexabundle is inset in the connector to place the hexabundle face at the focal plane of the telescope. A protective cylindrical plastic cap prevents impact damage on the hexabundle surface. There are 13 of these connectors, each housing one hexabundle, and each fits into pre-drilled holes in the field plate. The three main advantages of this new design are the inclusion of a magnet to simplify attachment to the plate, a ``key" on one side to orient the bundle in rotation, and a smaller footprint on the plate, which allows the hexabundles to be positioned closer to each other (within 15mm, equivalent to 228 arcsec). This enables more efficient tiling of galaxies in each field (see Section~\ref{sec_tiling}). Hexabundles are now positioned with an accuracy of less than half a fibre core (50$\mu$m, which is equivalent to $0.8''$; set by the machining tolerances) in linear position, and have a mean error of 0.55$^{
\circ}$ in rotation.

\subsection{Improved throughput}
\label{sec_thru}

The new fibre cables and hexabundles were tested for FRD and throughput before installation in SAMI-II.
Throughput was measured using the cut-back technique on 2 fibres in one slit
block. 2m of bare WF105/125 fibre was spliced to the $\sim40.7$m slit block
fibres and the throughput measured. The additional 2m was then cut off and
the throughput of the 2m alone was measured. This technique ensures that the cleaved
and mounted input end did not change, so that the throughput measures are
not affected by coupling into the fibre. Measurements were carried out through both
Bessel $B$ and $R$ filters in turn.
Table~\ref{thru} compares these throughput results to previous results from the
original SAMI AFS fibre, and from bare WF105/125 fibre. The new fibre run
with the slit block has a throughput that is similar to bare WF105/125 fibre, and clearly
much better than the AFS105/125 fibre used in the original SAMI instrument. In the blue, the fibre replacement gives a ~30\% gain in throughput for the fibre component of the SAMI system.

\begin{table}
\begin{center}
  \caption{Throughput of the upgraded SAMI-II fibre cable using WFS105/125 fibre with slit blocks attached, compared to both bare fibre of the same type (WF105/125), and to the previous SAMI fibre type (AFS105/125). Throughputs were measured through $B$ and $R$ Bessel filters, centred at 457nm (width 27nm) and 596nm (asymmetric profile of width 60 nm) respectively.
\label{thru}}
\begin{tabular}{lll}
\hline 
Fibre type  & \% Blue  & \% Red \\
(all 40+/-1m long) & throughput & throughput\\
\hline
AFS105/125 & 55 & 81\\
Bare WF105/125 & 83 & 91\\
WF105/125 fibre cable & & \\
with slit block & 82 & 91.5\\
\hline 
\end{tabular}
\end{center}
\end{table}

The original fibre cable for SAMI-I suffered from significant FRD, leading to losses of up to 50\% in the blue. This was due to the ribbonising of the fibres and the packaging method within the fibre cable \citep[see][for details]{Cro2012}. The new fibre cable was designed to minimise FRD by packing the fibres in groups of 21 within single furcation tubes. Each tube was less than half filled and each slit block of 63 fibres fed into 3 furcation tubes. In addition, the fibre cable into which these were packed was designed to minimise rotation and hence twisting of the fibres. The FRD of two of the new slit blocks with $\sim 40.7$\,m fibre cable attached, was tested before assembly of SAMI-II. Four fibre cores were tested in each slit block. The loss due to FRD in all four cores of the first slit block is $< 1$\%, while the other slit block measures FRD losses of up to 3.5\% in the Bessel $B$ filter band and 2.5\% in the Bessel $R$ filter band. The residual FRD is likely to be from compression of the fibres in the slit block glass  or compression/twisting of the fibres due to the memory effect of the short guiding tube that aligns the fibres into the slit block.

\begin{figure*}
\centerline{\psfig{file=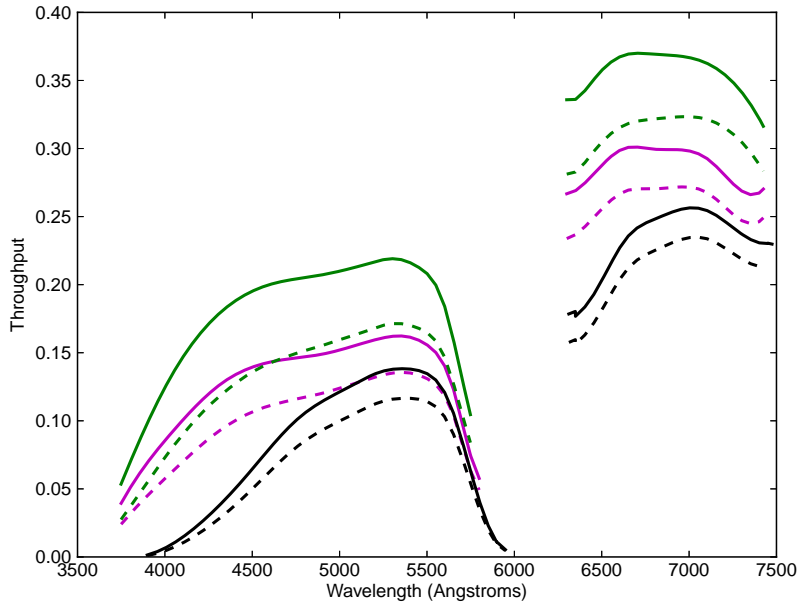, width=12.0cm}} 
\caption{Throughput of the SAMI instrument + AAOmega spectrograph + AAT telescope, excluding atmospheric losses (solid lines) and including losses from the atmosphere (dashed lines). Each line is based on standard stars observed in good conditions and is corrected for fill-fraction.  Stars observed with the original SAMI-I before the new hexabundles and fibre cable upgrade are shown in black, while those after the upgrade to SAMI-II before the CCD upgrade are in magenta. The green lines are for data taken with SAMI-II after the new blue AAOmega CCD was installed, the AAOmega optics were cleaned and the primary mirror was re-aluminised. 
The black curves sit below the magenta curves more towards the bluer wavelengths due to the difference in the fibre type and focal ratio degradation as reported in \citet{Cro2012} and discussed in Section~\ref{sec_thru}. The SAMI-I fibre (black) had lower blue throughput than the fibre used in the upgraded SAMI-II instrument (magenta).  The green lines have an improvement over the magenta curves that is consistent with the primary mirror increase in reflectivity of up to 17\% in addition to the cleaning of the optics which is estimated to have contributed of order 10\% in the blue and a little less in the red, and the upgrade of the blue CCD ($\sim$ 5\%).
}
\label{thrufig}
\end{figure*}

The total end-to-end throughput of SAMI was measured from standard star observations and in Figure~\ref{thrufig} we compare the throughput before the upgrade to that after the upgrade to SAMI-II. In each case several observations of a standard star taken in good conditions on a clear night were analysed and their throughputs were averaged. The throughput curves include all elements from the sky to the detector (telescope + spectrograph + SAMI) and are shown with and without the atmospheric losses. The two major improvements in SAMI throughput that are highlighted in this plot are firstly, the upgrade of the fibre cable and secondly, the new CCD and optics cleaning and we now discuss these in turn.

The original SAMI-I fibre cable (including the AFS105/127Y fibre) shows the lowest throughput, with a significant drop-off towards the blue. This drop-off is a combination of both the poor blue throughput of the original AFS105/125Y fibre and the FRD from the cable packaging and handling. 
At 4400\,\AA\ (centre of the Bessel $B$ filter from laboratory tests), the measured throughput is a factor of 1.8-2.2 lower than the theoretical value from fibre type alone. This agrees with the FRD laboratory tests of the SAMI-I cable as discussed in \citet{Cro2012}, in which the FRD resulted in a factor of up to $\sim 2$ lower throughput in the Bessel $B$ filter band. Similarly, in the blue end of the red (6400\,\AA) where the original cable was tested in the Bessel $R$ filter band, the losses had been a factor of $\sim 1.5$ in throughput due to FRD alone, which matches the improvement we now see with the new fibre cable.
Therefore we are confident that the new fibre type and cabling has removed the FRD losses and improved the fibre transmission as expected.

The second major improvement in Figure~\ref{thrufig} is highlighted by data taken after March 2014 (green). In March 2014, the primary mirror of the AAT was re-aluminised for the first time in several years, the blue CCD in AAOmega was upgraded \citep{Bro14} and the optics in AAOmega were thoroughly cleaned. The primary mirror reflectivity was measured to improve from $\sim75$\% to $\sim85-88$\% which is a factor of up to 17\% improvement. While the improvement from cleaning the AAOmega optics was not measured, it is estimated to be another 10\% improvement. The broad level of increase in throughput we measured in the data since then can be explained by this optics cleaning in the blue and the red. The expected improvement from the new CCD in the blue was around 5\%, which cannot be disentangled from the increased throughput due to the optics cleaning.
The AAOmega blue CCD upgrade also removed cosmetic defects, which assists in the SAMI data reduction and spectral line analysis.

\section{Science drivers influencing the target selection}
\label{TS}

The SAMI Galaxy Survey targets have been chosen to focus on a number of key science goals that depend primarily on a broad range of stellar mass and environment.
A detailed discussion of the science underpinning the SAMI Galaxy Survey is given in \citet{Cro2012}. Here we discuss each of the science selection requirements in turn that led to the final selection given in Sections~\ref{GAMA} and~\ref{Clusters}.

\subsection{Broad range in stellar mass}

The key science drivers for SAMI are dependent on a galaxy sample that is evenly distributed over a broad range of stellar masses. This requirement underpins the investigation of how both mass and angular momentum build in galaxies and how gas gets into and out of galaxies to regulate star formation. A range of stellar masses is also essential for consideration of how environment influences galaxy formation because lower stellar mass galaxies are thought to be more affected by their environment. 

Understanding galaxy formation requires reconciling theoretical cold dark matter (CDM) mass functions with observed stellar mass functions \citep{Bal08}. The deviation between them is most pronounced at both the lowest and highest stellar masses. Various feedback processes have been incorporated into the models to account for this, and therefore the feedback mechanisms need to be inherently mass-dependent. The link between feedback from winds or outflows and stellar mass, star formation rates, morphologies and the presence of AGNs is still inconclusive. Therefore testing different feedback processes requires an investigation across a large range of masses.

The Tully-Fisher Relation \citep[TFR;][]{Tull77} links two fundamental properties of disk galaxies:
their luminosity and their rotation velocity. This scaling (effectively between stellar mass and dynamical mass) underpins models of galaxy evolution, and becomes particularly interesting for $v < 100$\,km s$^{-1}$ where the velocity functions predicted by CDM and warm dark matter (WDM) models deviate significantly \citep{Zav09}. To investigate this parameter space the SAMI survey targets need to cover a sufficient radius to clearly define the circular velocity (as discussed further in Section~\ref{GAMA}). Furthermore, in order to reach galaxies with a nominal circular velocity of $\sim50$\,km s$^{-1}$ ($< 100$\,km s$^{-1}$, but large enough to be observable), the stellar mass selection must extend down to $\log_{10}(M_*/M_{\odot})\sim8.5$, based on the Tully-Fisher relation from \citet{Dut11} given by
\begin{equation}
\log_{10}\left(\frac{V_{2.2}}{{\rm km\,s^{-1}}}\right)=2.179+0.259\log_{10}\left(\frac{M_*}{10^{10.3}M_{\odot}}\right).
\end{equation} where $V_{2.2}$ is the rotation velocity measured at 2.2 disc scale lengths.

The tight relation between black hole mass and the galaxy's bulge mass measured from the velocity dispersion \citep{Tre02} implies a physical relation between the central black hole and star formation and hence the buildup of stellar mass.  Understanding accretion of gas and feeding of star formation requires a large stellar mass range because physical processes are different between higher and lower stellar mass galaxies \citep{Ker05}. For example, galaxies that have built a large stellar mass may have been fuelled by major interactions or events large enough to deposit sufficient gas to feed an AGN \citep{Vol03}. However, gas accretion in low stellar mass galaxies may be entirely different and due to the infall of non-shock heated gas or non-disruptive events \citep{Bro09}, yet still exhibit this same relation as those of very high stellar mass. 

Additionally, the shallow potential wells of low mass galaxies should make them more susceptible to the effects of feedback.  The mass loading of galaxy winds and wind velocity has been shown to scale with galaxy mass \citep{Mar05b,Hop12}.  Recent simulations also suggest that the transformation of cuspy dark matter density profiles into shallower cored profiles varies with mass \citep{Gov12,diC14}. Feedback also regulates star formation, and may be further hampered at the low metallicities
of dwarf galaxies \citep{Rob08,Gne10,Kru12}. Selecting a broad range in stellar mass allows an examination of how these processes scale with mass.

\subsection{Environment}

The key SAMI science drivers that require a broad range of environments include the mechanisms driving gas into and out of galaxies, and the impact of these gas flows on star formation. The environment influences star formation activity, galaxy colour and morphology. 

In denser environments processes such as ram-pressure stripping or strangulation can truncate star formation in the outer regions of a galaxy, or across the disk respectively \citep[e.g.][]{Bek09, Lew02} or induce star formation \citep[e.g.][]{Bek14}.  Furthermore, galaxy harassment \cite[e.g.][]{Moo98} or mergers and interactions between galaxies in groups and clusters can both strip gas and drive it towards the centre, triggering star formation \citep[e.g.][]{Ion04,Koo04}.  A range of environments are therefore necessary to understand the suppression of star formation and hence compare the outside-in to inside-out models \citep[e.g.][]{Cap13} of galaxy evolution. A dependence of star formation rate on local galaxy density has also been seen outside of clusters, in significantly less dense environments \citep{Mat03}. This dependence requires a physical mechanism which is not well understood due to the fact that studies outside clusters have mainly been based on fibre spectroscopy \cite[see also][]{Wij12}, while in clusters only a 2D approach can discriminate between various models. Small IFU studies \citep[e.g.][]{Bro13} have found no spatial dependence of star formation on environment however, a much larger sample is required.

Galaxies are known to evolve in colour (blue to red) and in morphology. In both cases there is a link to environment, with redder, and early-type galaxies predominantly found in higher density environments \citep{Dre80, Bes04}. However an understanding of these mechanisms requires IFU data for galaxies in a range of environments at each fixed stellar mass.

The kinematic morphology-density relation has shown that slow rotators are preferentially found in the densest regions, while there is a transition from spirals to early-type fast-rotators with increasing density \citep{Cap13}. An understanding of this relationship requires addressing fundamental questions about how slow- and fast-rotators are formed, including the effects of minor and major mergers and environmental density. It is unclear whether slow-rotators form only in dense regions or migrate there. The SAMI survey will amass one of the largest samples of spatially-resolved early-type galaxies to date, and will have a broad range of environments from the dense cluster regions to field galaxies to clearly define the kinematic-morphology density relation \citep{Fog2014}.

\section{Science constraints on the target selection to meet the science drivers}
\label{constraints}

The multiplex of the SAMI instrument corresponds to 12 galaxies in each 0.79 square degree field. Therefore optimal observational efficiency requires a suitable balance between redshift range and stellar mass cuts. We selected a redshift range and then set the stellar mass limits to give the broadest mass range. The considerations in the following discussion influenced these boundaries.

\subsection{Redshift range}

\begin{figure}
\includegraphics[width=60mm,angle=270]{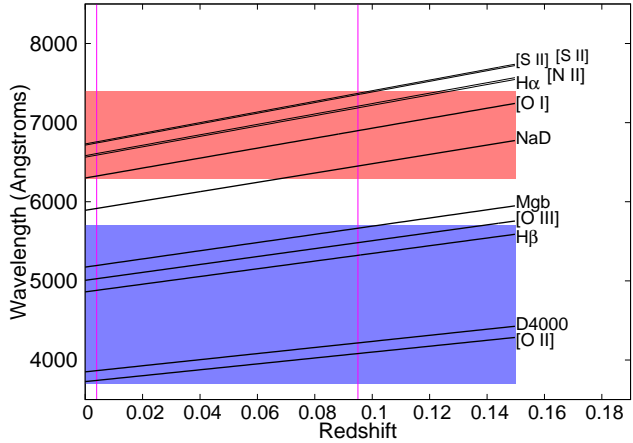}
\caption{Wavelength coverage vs redshift for the current default SAMI
  configuration compared to key spectral lines. The red and blue regions indicate the wavelength coverage of the red and blue arms of AAOmega. Magenta lines mark the redshift boundaries for the primary sample. 
 }
\label{speclines}
\end{figure}

The SAMI instrument feeds into the AAT's AAOmega spectrograph \citep{Sha06}. While this is a versatile spectrograph, the SAMI survey adopts a fixed resolution and wavelength range for AAOmega. Using the 580V grating in the blue arm of the spectrograph gives a resolution $R\sim 1700$, while the 1000R grating in the red arm results in $R\sim 4500$. The wavelength range covered is 3700--5700\,\AA\ in the blue and 6300--7400\,\AA\ in the red. This setup was adopted to optimise coverage of important spectral features in the ideal redshift range while maintaining high spectral resolution in the red for kinematic analysis, as shown in Figure~\ref{speclines}. Table~\ref{zlines} shows the redshift ranges afforded by this wavelength range for the key lines, allowing
for a 20\,\AA\ window at either side of the line. Redshift limits for the SAMI Galaxy Survey were restricted to $z < 0.095$ so that \sii\ $\lambda\lambda$\,6716, 6731 and  Mgb\,$\lambda$\,5179 in most cases will be within the in the red and blue bands respectively. 
A new dichroic is planned for the spectrograph which will extend the blue arm coverage towards the red.

\begin{table}
\begin{center}
\caption{Redshift coverage for each spectral feature within the fixed SAMI Galaxy Survey resolution and wavelength range.  }
\label{zlines}
\begin{tabular}{llcccc}
\hline 
 Line  & $\lambda_{rest}$ & $z_{min}(B)$ &  $z_{max}(B)$ & $z_{min}(R)$ &  $z_{max}(R)$  \\
\hline
\oii     & 3727 &   0.004   &0.521 & 0.699$^{*}$  &0.975$^{*}$ \\
D4000 & 3850  &  0.004  & 0.473 & 0.644$^{*}$  &0.912$^{*}$ \\
\hb       & 4861   & 0.004  & 0.168 & 0.301$^{*}$  &0.516$^{*}$ \\
\oiii    & 5007  &  0.004  & 0.134 & 0.263$^{*}$  &0.472$^{*}$ \\
Mgb    & 5174   & 0.004  & 0.097 & 0.222$^{*}$  &0.425$^{*}$   \\
NaD     & 5892 &$-$ &$-$ & 0.073  &0.251\\
\oi      & 6300 & $-$ &$-$ & 0.004  &0.171\\
\ha       & 6563 & $-$ &$-$ & 0.004  &0.124\\
\nii     & 6583 & $-$ &$-$ & 0.004  &0.121\\
\sii      & 6716 & $-$ &$-$ & 0.004  &0.099\\
\sii      & 6731 & $-$ &$-$ & 0.004  &0.096\\
\hline 
\end{tabular}
\end{center}
$^{*}$Values with redshift too high to be used by the SAMI Galaxy survey.
\end{table}

The redshift range also sets the spatial resolution achieved within a SAMI hexabundle. Our redshift range means that one fibre core images 0.1--2.8\,kpc, and a hexabundle diameter is 1.2--26.6\,kpc. We have not made any selection based on major axis effective radius, $R_e$, and the spatial distribution that results from our sample is discussed further in Section~\ref{size_sb}.

\subsection{Stellar mass or absolute magnitude selection}

It is important to have well defined boundaries on the stellar mass selection, to facilitate accurate volume corrections. Therefore careful consideration was given to whether the selection limits should be based on  stellar masses or absolute magnitudes, and a summary of the key issues is given in Table~\ref{tradeoffs}. 

We adopted a combined approach by using absolute magnitudes and colours to calculate a proxy for stellar mass which was then used as the basis for the selection. This choice was primarily driven by the well defined limiting values for absolute magnitudes, that are not model-dependent, and therefore ensure the survey limits will not change or become dispersed. The method follows \citet[][eq. 8]{Tay11}, which showed that the GAMA survey stellar masses generated by optical SED fitting could be approximated by 
\begin{equation}
\log_{10}(M_*/M_{\odot})=1.15 + 0.70(g-i)_{{\rm rest}} - 0.4M_i,
\label{Eq_Tay}
\end{equation}
where $M_i$ is the AB rest-frame $i$-band absolute magnitude, and $M_*$ is the stellar mass in solar units.
Applying Equation~\ref{Eq_Tay} to the SAMI survey, we used observed-frame Milky-Way-extinction-corrected apparent magnitudes ($g$ and $i$), and limited the colours to reasonable values of $-0.2 < g-i < 1.6$. We calculated the stellar mass using
\begin{multline}
\log_{10}(M_*/M_{\odot})=-0.4i + 0.4D - \log_{10}(1.0+z)\\+(1.2117-0.5893z)+(0.7106-0.1467z) \times (g-i)
\label{SM_eq}
\end{multline}
where D is the distance modulus.
We used the aperture-matched $g-$ and $i-$band photometry from the GAMA catalogue  \citep[`auto' magnitudes;][Liske et al. in prep.]{Hil11} in the regions of overlap, and aperture-matched photometry generated from the SDSS or the VLT Survey Telescope (VST) ATLAS imaging data \citep[VST/ATLAS; http://astro.dur.ac.uk/Cosmology/vstatlas/;][]{Sha13} in the remaining cluster regions (for details see Owers et al., in prep).  
This selection method allows for accurate volume corrections.

\begin{table*}
\begin{center}
  \caption{Trade-off between stellar mass and absolute magnitude selection.  }
\label{tradeoffs}
\begin{tabular}{| r | p{7cm} | p{7cm}}
\hline 
 & \multicolumn{2}{c}{Selection basis}\\
 \cline{2-3}
  & Stellar mass & Absolute magnitude \\
\hline
Advantages & \begin{itemize}
    \item Simplifies comparison of properties as a function of stellar mass. 
    \item Stellar mass estimates are based on model fitting and allow for dust obscuration. 
\item Changes to measurements of dust, metallicity and galaxy ages will not significantly alter the stellar mass values, as these quantities are correlated.
\item The broad range of stellar masses required and distribution of stellar masses is simple to select.
\end{itemize} &
\begin{itemize}
\item Not dependent on modelling, other than the k-correction, which is an interpolation of
  the multi-band photometry, and thus relatively insensitive to systematics.
\item The effective volume of the survey can be well defined.
\item Future changes to models will not significantly affect the selection boundaries.
\end{itemize} \\ \hline
Disadvantages & \begin{itemize}
\item  Stellar mass estimates are derived from stellar population synthesis fitting \citep[e.g.][]{Bru93}, which uses stellar evolution models \citep[e.g.][]{Bru03}, star formation histories and the assumption of a stellar initial mass function (IMF). The stellar masses will therefore change if the IMF or stellar evolution model is updated. This would lead to a poorly defined boundary to the source selection and an ill-defined volume. Based on the magnitude of systematic IMF variations as a function of M/L reported by the ATLAS$^{\rm 3D}$ team  \citep{Cap13}, we estimate that 
variations in IMF will give
an error of order $\sim0.3$ dex in the stellar masses used to define the sample.
\end{itemize} &

\begin{itemize}
\item Absolute magnitude is not a direct measure of stellar mass. However, stellar mass can be found by combining the absolute magnitude with rest-frame colour \citep{Tay11}. 
\item The scatter in the \citet{Tay11} correlation between galaxy colours and stellar mass means that galaxies at the limits of our selection will favour slightly bluer galaxy types.
\end{itemize} \\ 
\hline 
\end{tabular}
\end{center}
\end{table*}

\section{Final selection of non-cluster galaxies}
\label{GAMA}

The galaxy survey regions have been selected to be the equatorial G09, G12 and G15 regions from the GAMA galaxy redshift survey, whose coordinates are given in Table~\ref{GAMAcoords}.

 \begin{table}
\begin{center}
\caption{Coordinates of the GAMA-I fields used in the SAMI survey. Each region is $12^{\circ}\times 4^{\circ}$.
\label{GAMAcoords}}
\begin{tabular}{lcc}
\hline 
Field & R.A. (J2000) & Dec. (J2000)  \\
 & ($^{\circ}$) & ($^{\circ}$) \\
\hline
G09 & 129.0 -- 141.0 & -1.0 -- +3.0 \\
G12 & 174.0 -- 186.0 & -2.0 -- +2.0 \\
G15 & 211.5 -- 223.5 & -2.0 -- +2.0 \\
\hline 
\end{tabular}
\end{center}
\end{table}

\subsection{The GAMA survey}

The GAMA project brings together multi-wavelength data from the far-UV to infrared, and is centred around a single-fibre galaxy redshift survey \citep{Hop13} that was observed from 2008-2014 using the 2dF instrument with the AAOmega spectrograph on the AAT. The photometric $ugriz$ optical data and positions from the GAMA input catalogues are drawn
directly from the SDSS Data Release 7 \citep{Aba09}.
GAMA-I (the first stage of GAMA adopted by SAMI) therefore provides magnitudes and redshifts in three equatorial fields centred at $9^{h}00^{m}$ +$1^{d}$ (G09), $12^{h}00^{m}$ +$0^{d}$ (G12) and $14^{h}30^{m}$ +$0^{d}$ (G15). Each field is $12^{\circ} \times 4^{\circ}$, giving 144 square degrees in total. 
The GAMA targets were selected from SDSS based on the Petrosian\footnote{Magnitudes measured in a circular aperture that has twice the Petrosian radius determined from $r$-band surface brightness.} and model\footnote{Magnitudes based on the best fit to an exponential or de Vaucouleurs profile.} magnitudes, with $r_{petro} < 19.8$, or $z_{model} < 18.2$ and $r_{model} < 20.5$ or K\footnote{AB magnitude using an elliptical aperture based on the \citet{Kro80} algorithm.}$_{auto} < 17.6$ and $r_{model} < 20.5$ in all three regions. The redshifts are generally $z < 0.5$ with a median of 0.2. The GAMA-I redshift completeness is $\geq98$\%. GAMA is therefore ideal for selecting SAMI galaxy field and group targets. Full details of the GAMA target selection can be found in \citet{Bal10}, and the GAMA data release 2, which is now mostly public, is described in Liske et al. (in prep.).

A key motivation for choosing the GAMA regions is the supporting data in the ultraviolet (UV), near- and far-infrared (IR) and at radio wavelengths, as detailed in Driver et al. (in prep.). This ancillary data will add further value to the SAMI observations and is discussed in Section~\ref{ancil}.

\subsection{Limits for the stellar mass selection}

The aim is to select a broad range in stellar mass. This could be done with either a cut-off defined by a single smoothly-varying function in the proxy for stellar mass with redshift, or alternatively a stepped series of stellar mass limits that change with redshift.  
A single selection function means that volume limited samples could not be considered without disregarding many of the galaxies observed, or applying a weighting function to do volume corrections \citep[as done in][]{Wal14}. However, a stepped series of stellar mass limits forms a number of separate volume-limited samples.

While we aim for a uniform distribution of stellar mass, the GAMA-I sample from which we are selecting SAMI galaxies contains clear density structures (see Figure~\ref{Mstar}) and we do not want to bias against proper sampling of these structures. Therefore the final stellar mass distribution will necessarily not be entirely flat. 
Figure~\ref{Mstar} illustrates a selection in which there are 150 objects in each 0.25 dex bin (blue  lines joins bin centres) as well as the limits on 1 dex bins with 540 galaxies in each (green line). This result indicates that a strictly uniform distribution in stellar mass with a single function would only be achievable with a non-linear selection function (as shown) in redshift versus stellar mass space. A uniform stellar mass selection would result in $\sim 0.5$ dex change in stellar mass across an individual structure (e.g. filament, group) that spans $cz = 1000\,$km\,s$^{-1}$ in the steepest part of the function. For example, the green and blue lines cut through the large structure at $\sim0.05 < z < 0.06$, while the final selection (red line) was shifted so the structure fit within a single stellar mass bin. It is far preferable to have the same stellar mass limits for galaxies within bound structures.

Therefore, in order to remove selection biases and simplify volume corrections, we have chosen a stepped function defining several volume-limited samples with semi-regular redshift intervals and mass steps, at the expense of strict uniformity in the stellar mass distribution.

\subsection{Selection of SAMI targets from the GAMA-I survey}
\label{finalGAMA}

The SAMI sample is drawn from the GAMA data set, combining several GAMA catalogues\footnote{The GAMA catalogues used here are TilingCatv29.fits \citep{Bal10}, ApMatchedCatv03.fits \citep{Hil11}, GalacticExtinctionv02.fits, DistancesFramesv08.fits \citep{Bal12}, SersicCatAllv07.fits \citep{Kel12}, StellarMassesv08.fits \citep{Tay11}, EnvironmentMeasuresv01.fits \citep{Bro13} and InputCatAv06.fits \citep{Bal10}, and they can be found on the GAMA webpages.}.
From this combined catalogue we selected the SAMI galaxies by firstly only including the objects within the GAMA-I regions, and secondly, rejecting objects with unreliable redshifts (quality flag nQ$\leq2$), or unreliable magnitudes ($g$, $r$ or $i$ auto mags $< 0$ or $> 90$). The resultant SAMI field catalogue in the GAMA regions includes the data types listed in Table~\ref{catdata}. 

\begin{figure*}
\centerline{\psfig{file=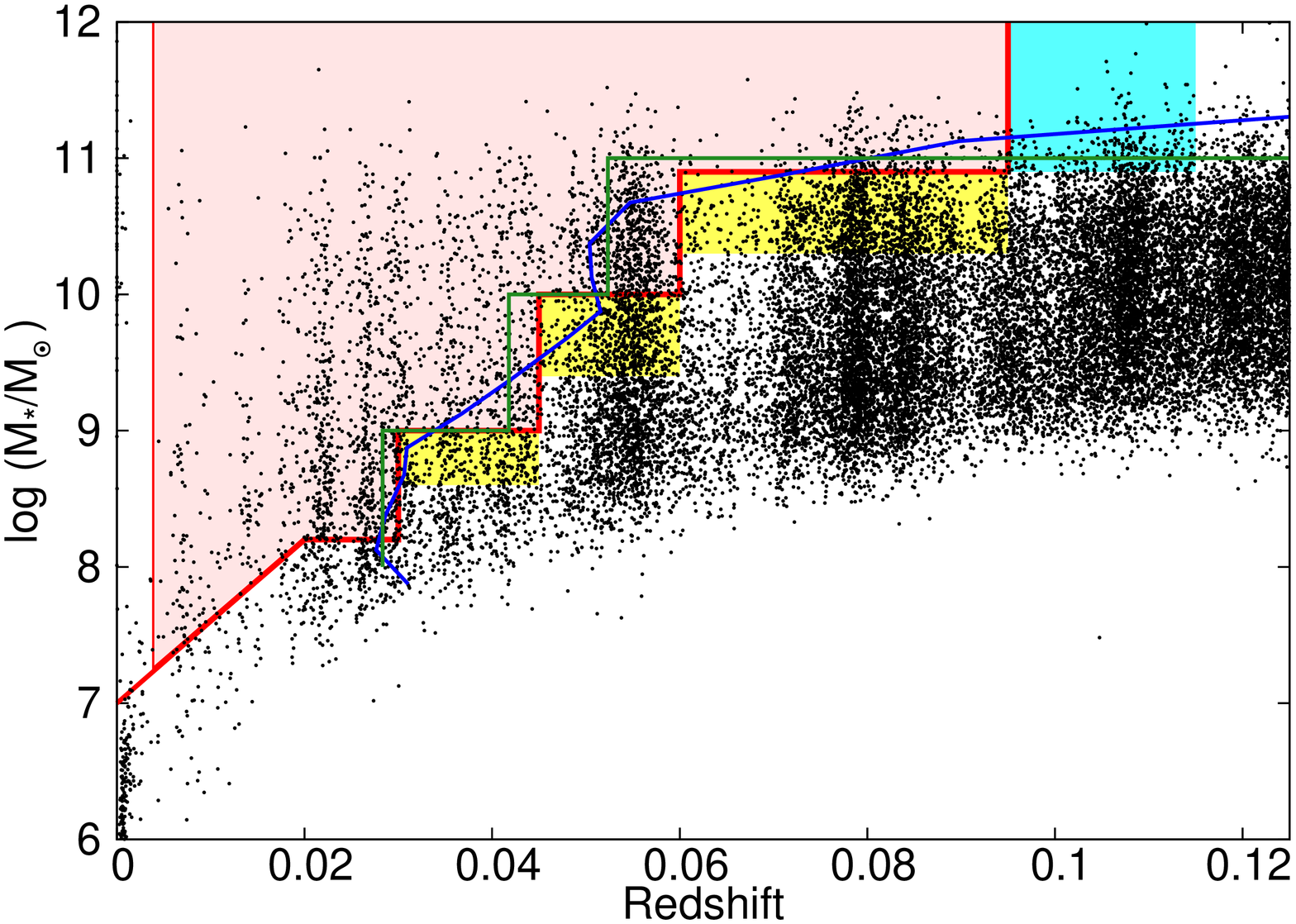, width=10.0cm, angle=-90}}
\caption{Stellar mass and Tonry redshift (adjusted to the \citet{Ton00} flow model) distribution defining the selection of SAMI galaxies from the GAMA-I catalogue. Black points are the GAMA catalogue from which SAMI targets were selected. In the final selection, the highest priority targets lie above the red line and within the redshift range $z = 0.004$ to 0.095 (pink region), while the yellow and cyan boxes represent lower priority targets to be used as fillers in pointings where 12 high priority targets cannot be optimally tiled within the $1^{\circ}$ diameter field. The blue and green lines are not the selection boundaries; they just indicate a flat stellar mass distribution for comparison. The blue line marks bin centres for where a continuous selection would lie in order to give 150 galaxies in each 0.25 dex stellar mass bin, while the green line marks the limits of 1 dex stellar mass bins that have 540 galaxies each. 
}
\label{Mstar}
\end{figure*}

The SAMI galaxies selected from the GAMA survey consist of four volume-limited samples from a stepped series of stellar mass cuts in redshift bands as shown in Figure~\ref{Mstar}, along with additional dwarf galaxy candidates with low stellar mass and low redshift.
The points above the red line and within the redshift range $z = 0.004$ to 0.095 (pink region) define the main sample which has limits set as described in Table~\ref{limits}.  Due to tiling constraints and source distributions some field configurations may not have 12 primary targets, therefore filler targets were also defined.  In the yellow regions (with lowered cut-offs of log$(M_*/M_{\odot}) = 8.6, 9.4$ and 10.3) we have additional lower priority targets to use as fillers. A selection of higher redshift galaxies (cyan region) are further filler targets with $0.095 < z < 0.115$ and log$(M_*/M_{\odot}) > 10.9$. 

\begin{table*}
\begin{center}
  \caption{Data included in the final SAMI GAMA-region catalogue. }
\label{catdata}
\begin{tabular}{llllll}
\hline 
Name & Index  & Units & Description & Source GAMA-I Catalogue\\
\hline
NAME                         &  1            &     & IAU format object name&\\
RA	                          &  2		&degrees	&J2000 coordinate& TilingCatv29  \\  
DEC	                          &  3		&degrees	&J2000 coordinate& TilingCatv29   \\  
r\_petro          &  4		&mag&	Extinction-corrected SDSS DR7 Petrosian mag& TilingCatv29  \\  
r\_auto	  &  5		&mag&	Extinction-corrected Kron magnitude (r band)& ApMatchedCatv03 MAG\_AUTO\_R  \\  
z\_tonry   	          &  6		&	&Flow-corrected redshift using Tonry  model& DistancesFramesv08   \\  
z\_spec     		& 7 &  & Spectroscopic redshift & \\
M\_r	          &  8		&mag&	Absolute magnitude in restframe $r$-band &StellarMassesv08 absmag\_r \\
  &  &  & from SED fits&   \\  
r\_e	          &  9 		&arcsec&	Effective radius in $r$-band (hl rad) (semi-major)& SersicCatAllv07 GAL\_RE\_R   \\  
$<$mu(re)$>$ &  10 		&mag\,arcsec$^{-2}$&	Effective $r$-band surface brightness within $r_{e}$& SersicCatAllv07   GAL\_MU\_E\_AVG\_R \\  
mu(re)	  &  11		&mag\,arcsec$^{-2}$&	Effective $r$-band surface brightness at $r_{e}$&  SersicCatAllv07  GAL\_MU\_E\_R \\  
mu(2re)	  &  12		&mag\,arcsec$^{-2}$&	Effective $r$-band surface brightness at 2$r_{e}$&   SersicCatAllv07  GAL\_MU\_E\_2R \\  
ellipticity & 13 &  &Ellipticity from $r$-band Sersic fits & SersicCatAllv07 GAL\_ELLIP\_R\\
PA & 14 & degrees & Position angle from $r$-band Sersic fits & SersicCatAllv07 GAL\_PA\_R\\
log$(M_*/M_{\odot})$	                  &  15		&dex&	Stellar mass based on Eq.~\ref{SM_eq}, & \\
	                  &  	&&$\log(M_*/M_{\odot})$&      \\  
$g-i$	  &  16		&mag&	Kron colour, extinction corrected &  derived from ApMatchedCatv03  MAG\_AUTO\_G-I \\  
A\_g	                          &  17		&mag&	Galactic extinction in SDSS g band&  GalacticExtinctionv02  \\  
CATAID	                  &  18	&	& GAMA ID in InputCatAv06& TilingCatv29     \\  
SURV\_SAMI                 &  19    & & \multicolumn{2}{l}{Sample priority class:}\\
	&	&	&\multicolumn{2}{l}{primary sample (red region in Figure~\ref{Mstar}) = 8; }\\
	&	&	&\multicolumn{2}{l}{high-mass fillers (cyan region in Figure~\ref{Mstar}) = 4;}\\
	&	&	&\multicolumn{2}{l}{remaining fillers (yellow regions in Figure~\ref{Mstar}) = 3;}\\
PRI\_SAMI                    &  20     & & \multicolumn{2}{l}{Sample priority used for tiling galaxies. }\\
	&	&	&\multicolumn{2}{l}{Values are the same as for SURV\_SAMI except objects}\\
	&	&	&\multicolumn{2}{l}{have PRI\_SAMI set to 1 if BAD\_CLASS = 1--4 or 6--7,}\\
	&	&	&\multicolumn{2}{l}{ or once they have been observed (i.e. OBS\_SAMI = 1).} \\
BAD\_CLASS                & 21      & & \multicolumn{2}{l}{Classification based on visual inspection (see Section~\ref{finalGAMA}). }\\
	&	&	&\multicolumn{2}{l}{0 = object is OK; }\\
	&	&	&\multicolumn{2}{l}{1 = nearby bright star; }\\
	&	&	&\multicolumn{2}{l}{2 = target is a star; }\\
	&	&	&\multicolumn{2}{l}{3 = subcomponent of a galaxy; }\\
	&	&	&\multicolumn{2}{l}{4 = very large, low redshift galaxy;}\\
	&	&	&\multicolumn{2}{l}{5 = needs re-centring; }\\
	&	&	&\multicolumn{2}{l}{6 = poor redshift; }\\
	&	&	&\multicolumn{2}{l}{7 = other problems, }\\
	&	&	&\multicolumn{2}{l}{8 = smaller component of a close pair of galaxies, }\\
	&	&	&\multicolumn{2}{l}{where the second galaxy is outside of the bundle radius.}\\
	&	&	&\multicolumn{2}{l}{Only objects with BAD\_CLASS = 0, 5 or 8 will be in the sample that may be observed. }\\
OBS\_SAMI                  & 22 & & \multicolumn{2}{l}{Flag if object has been observed by the SAMI survey; 1 = yes, 0 = no} \\
TILE\_NUM & 23 & &  \multicolumn{2}{l}{Tile number or numbers that the galaxy was observed on, if it has been observed}\\
\hline 
\end{tabular}
\end{center}
\end{table*}

\begin{table}
\begin{center}
\caption{Selection boundaries in Tonry redshift (adjusted to the \citet{Ton00} flow model) and stellar mass for the primary and filler SAMI targets selected from GAMA.
\label{limits}}
\begin{tabular}{lll}
\hline 
Redshift & log$(M_*/M_{\odot})$ & Figure~\ref{Mstar}\\
 range & & colour \\
\hline
\multicolumn{3}{l}{Primary targets} \\
$0.004 < z \leq 0.02$ &$\geq 7 + 60z$& pink\\
$0.02 < z \leq 0.03$ & $\geq 8.2$& pink\\
$0.03 < z \leq 0.045$ & $\geq 9.0$& pink\\
$0.045 < z \leq 0.06$ & $\geq 10.0$& pink\\
$0.06 < z \leq 0.095$ & $\geq 10.9$& pink\\
\multicolumn{3}{l}{Filler targets}\\
$0.03 < z \leq 0.045$ & $\geq 8.6$& yellow\\
$0.045 < z \leq 0.06$ & $\geq 9.4$& yellow\\
$0.06 < z \leq 0.095$ & $\geq 10.3$& yellow\\
$0.095 < z \leq 0.115$ & $\geq 10.9$& cyan\\

\hline 
\end{tabular}
\end{center}
\end{table}

The total number of galaxies included in this selection is 2738 main survey targets and 2798 filler targets. The volume surveyed within the primary sample in the three GAMA regions is $3.18 \times 10^5 h^{-3}$\,Mpc$^3$ for Hubble parameter $h$. However this selection was refined further by visual inspection. 
The visual classifications used, and which classes were included and excluded, is given in Table~\ref{class}. Eleven percent of the objects were removed from the sample due to this classification, leaving 2404 main and 2513 filler targets.  
This selection on average satisfies the density of targets required to fill 12 hexabundles with primary targets in a 0.79 square degree field, with 14, 15 and 21 objects/degree$^{2}$ in the 9, 12 and 14.5 hour regions or 11, 12 and 16 objects per SAMI field of view respectively.

\begin{table*}
\begin{center}
\caption{Visual confirmation of SAMI galaxies, and the fraction that were removed from the sample for listed reasons.
\label{class}}
\begin{tabular}{lll}
\hline 
\% GAMA galaxies  & Observe & Classification \\
\hline
88.2   &    Y  & Confirmed to have no problems. \\
8.6 &  N  &  Bright star nearby. \\
0.1  &  N &  Target is a star. \\
1.4 &   N & Subcomponent of a galaxy, or interacting companion where the companion\\
     &        & is within 1 bundle radius. \\
0.3 &  N & Target is a very low redshift, very large, bright galaxy.  \\
0.6 &  Y & Position needs to be manually adjusted to centre targets in hexabundle.\\
0.8 &    N &  Catalogue redshift is in error. \\
\hline 
\end{tabular}
\end{center}
\end{table*}

\begin{figure*}
\centerline{\psfig{file=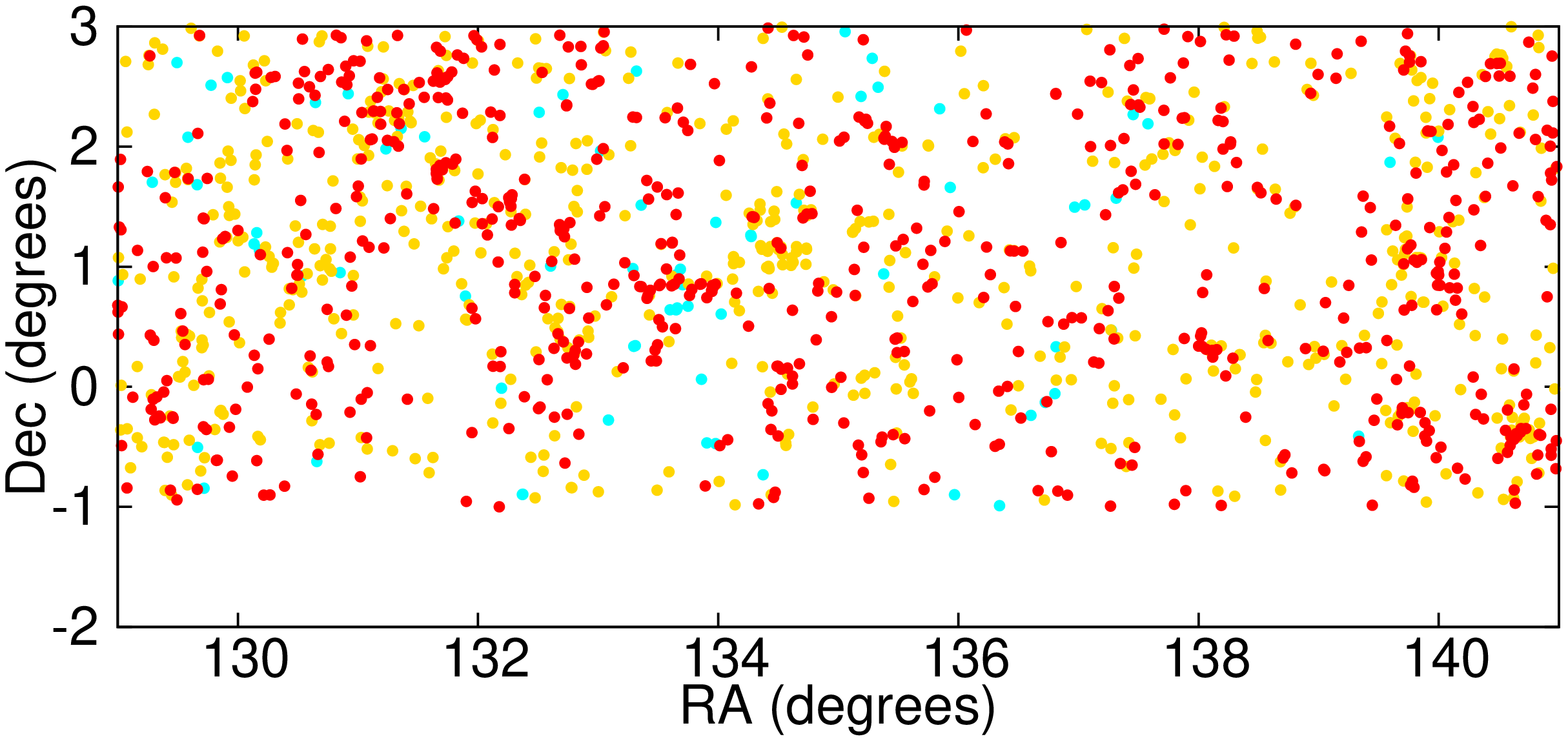, width=4.0cm, angle=-90}}
\centerline{\psfig{file=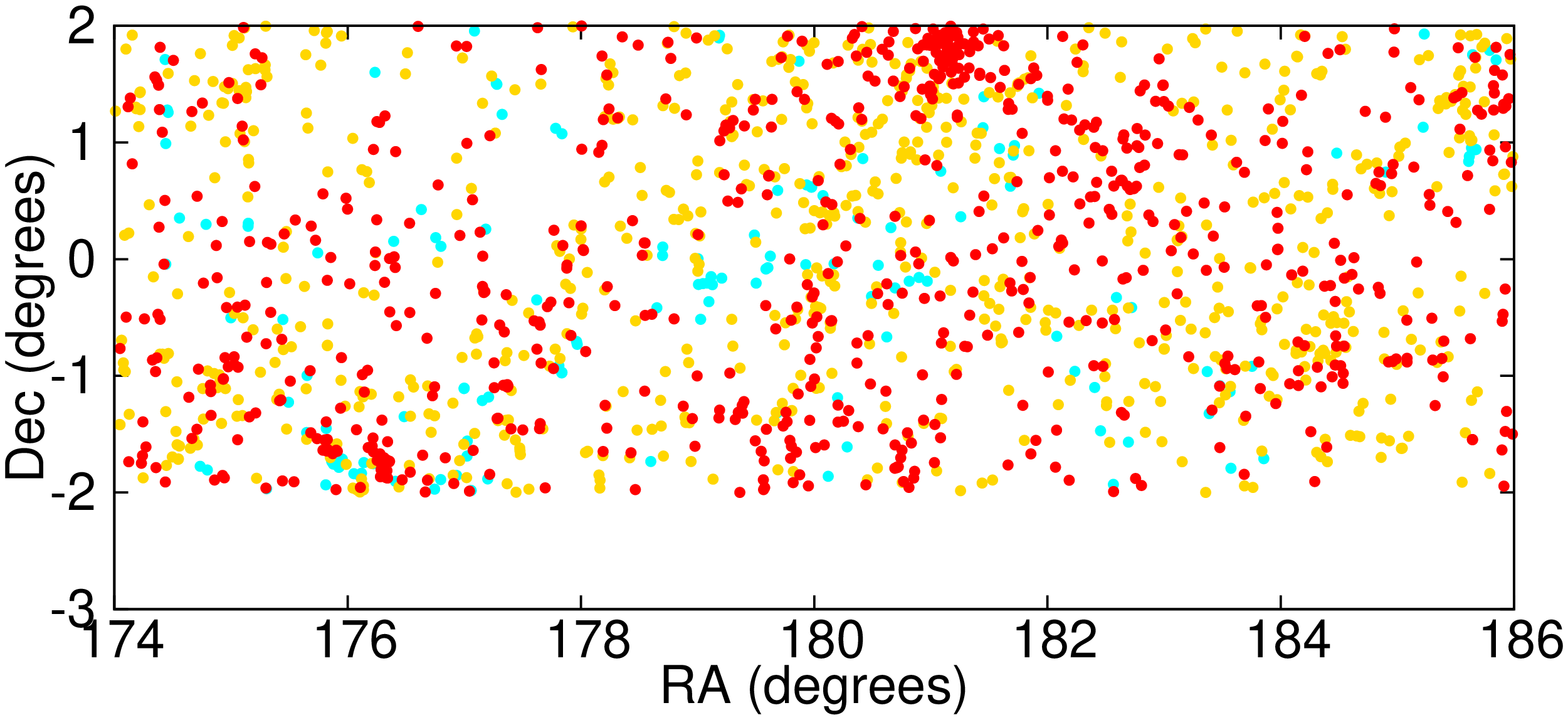, width=4.0cm, angle=-90}}
\centerline{\psfig{file=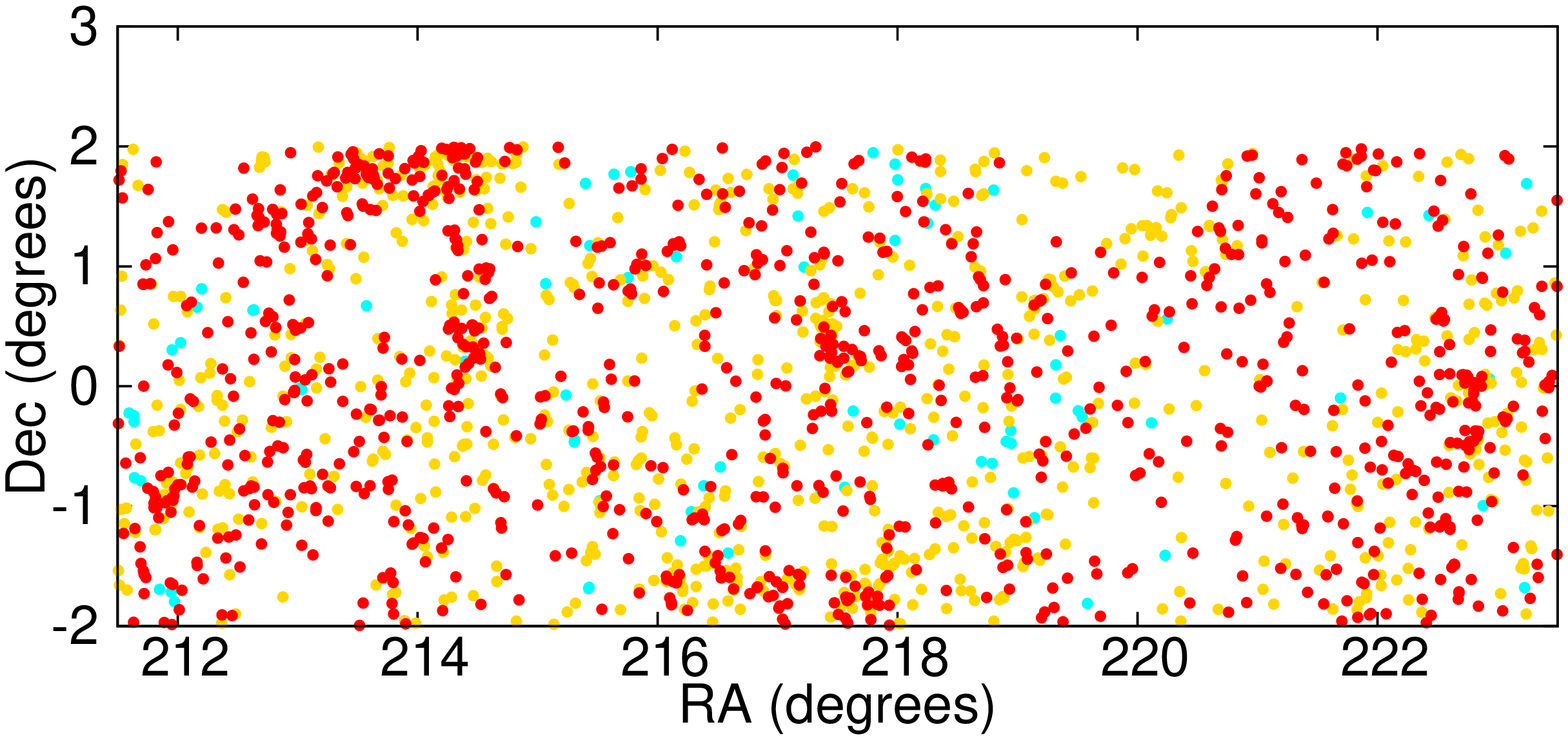, width=4.0cm, angle=-90}}
\vspace*{3mm}
\centerline{\psfig{file=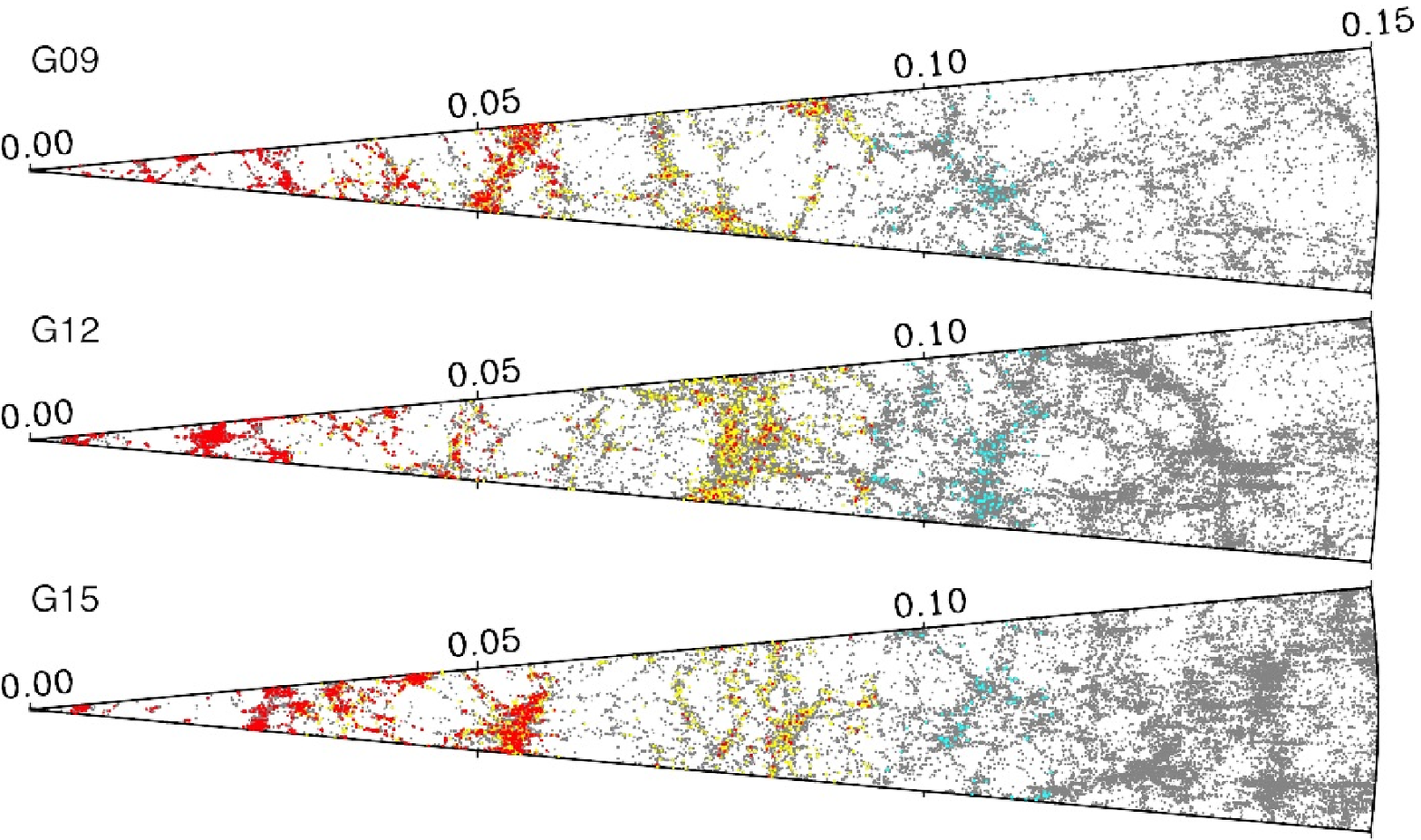, width=12.0cm}}
\vspace*{3mm}
\caption{On-sky distribution for SAMI targets within the G09, G12 and G15 regions (top three plots respectively). Redshift cone diagrams (bottom three plots) of the GAMA galaxies (grey) and the SAMI  Galaxy Survey targets. Colours of the SAMI galaxies correspond to regions in Figure~\ref{Mstar}, with the main sample in red, and the two filler sample galaxies in yellow and cyan.
}
\label{RA_DEC_z}
\end{figure*}

Figure~\ref{RA_DEC_z} shows the on-sky and redshift distribution of the SAMI galaxies, highlighting the large-scale structures traced by the SAMI survey. Figure~\ref{Mstar} shows that while the galaxy masses peak around $M_{\star}$ due to this large scale structure,  stellar masses from $10^{8}$ to $10^{11.5}\,M_{\odot}$ are well sampled. The parent population from GAMA has a slightly higher mass peak between $10^{10.5}$ to $10^{11}\,M_{\odot}$ \citep{Tay11}, which is similar to the peak of the mass distribution from the CALIFA survey \citep{Wal14} despite the difference in redshift range.

Limits on stellar mass at each redshift are set to make sure we attain the S/N required in the continuum or emission lines to achieve some science objectives in even the low surface brightness targets, while at the same time ensuring that the highest redshift galaxies have sufficient resolution elements. This design is tested with current data in Section~\ref{quality}.

We cannot  by design select galaxies based on environment as there is not a single well-defined environmental metric.  Rather we have tested the environments covered by our stellar mass selection to ensure a broad range. Galaxies that have been chosen from the GAMA fields are predominantly field galaxies and groups (see Figure~\ref{groups} later), and these have been supplemented by several galaxy clusters (see Section~\ref{Clusters}), in order to increase the sample of galaxies in higher-density environments and extend the range of environments covered.

\subsection{Galaxy sizes and surface brightness distributions}
\label{size_sb}

The range of galaxy sizes compared to the hexabundle size is crucial to the science of the survey. On the one hand, if the hexabundle radius samples $\leq1 R_e$ then the central stellar and gas distributions and dynamics can be investigated. However, in order to measure global dynamics (e.g. for the Tully Fisher relation) it is preferable for the hexabundle radius to extend to $> 2R_e$.

\begin{figure*}
\centerline{\psfig{file=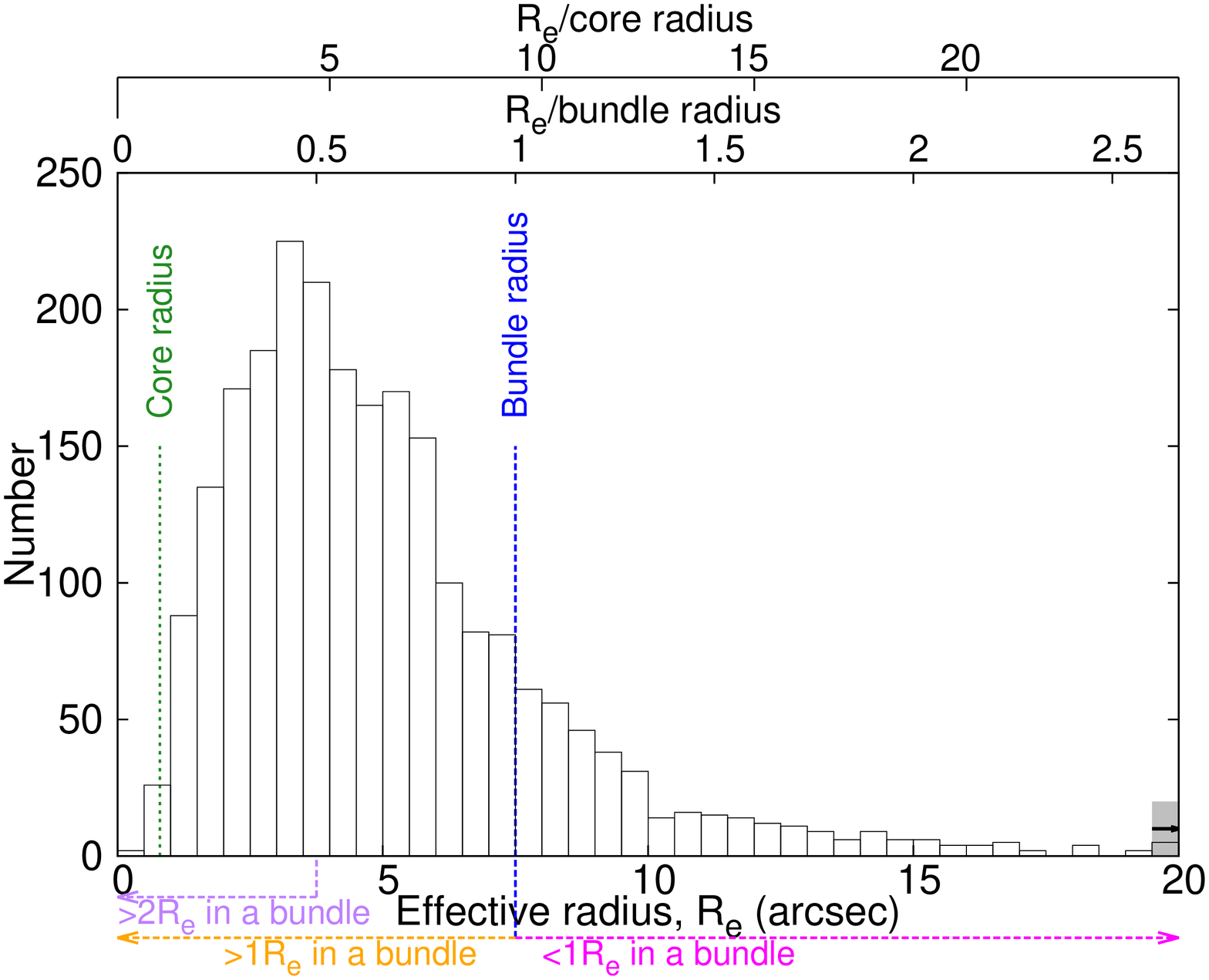, width=8.0cm, angle=-90}}
\vspace*{6mm}
\centerline{\psfig{file=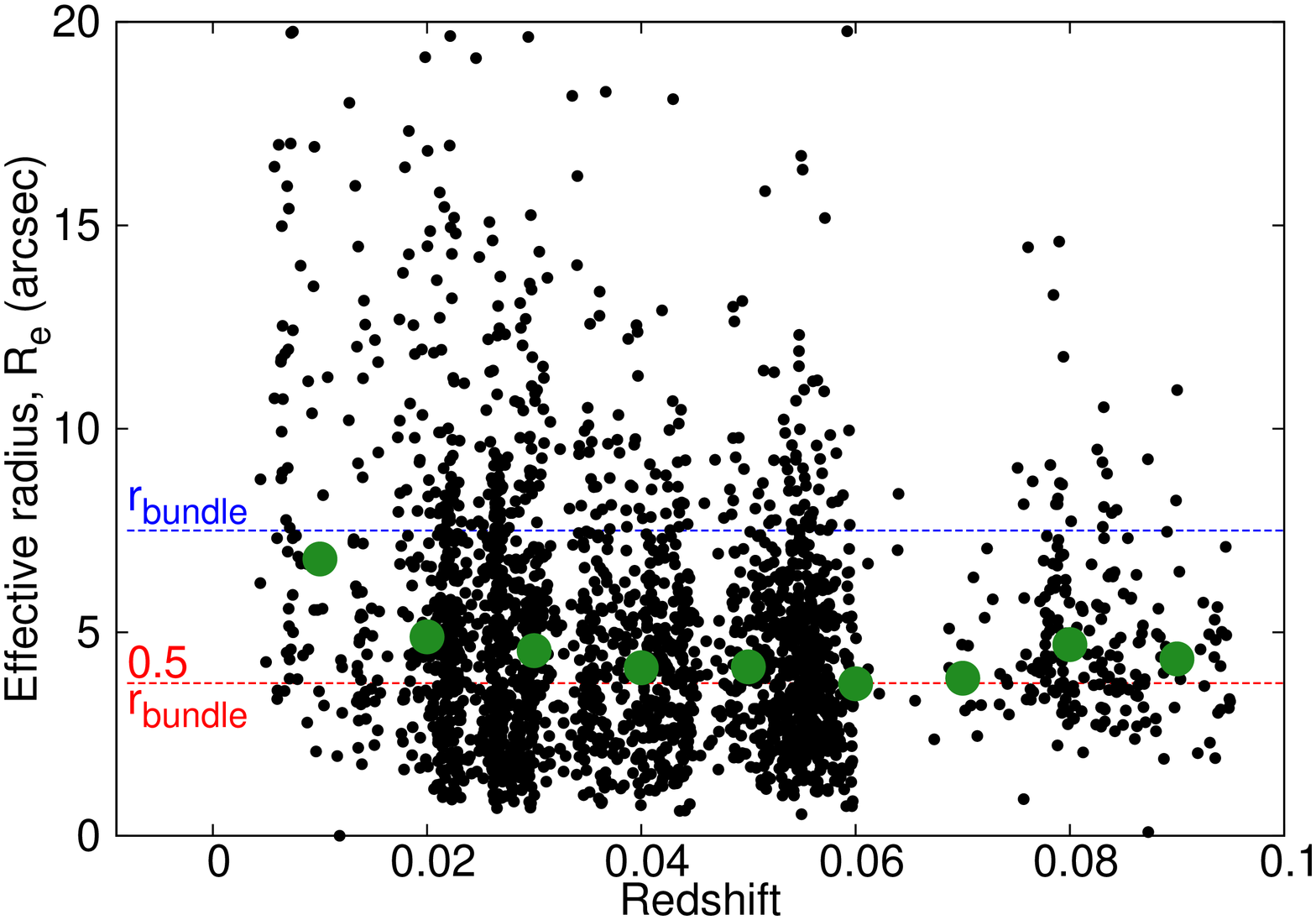, width=8.0cm, angle=-90}}
\caption{Top: The distribution of major axis effective radii for the primary SAMI targets, as measured from Sersic fits from the GAMA data (see Table~\ref{catdata}). The SAMI hexabundles each have a radius of 7.5 arcsec (blue line), while each of the 61 individual fibre cores has a radius of 0.8 arcsec (green line). Arrows mark the range of galaxies for which there are $> 2R_e$, $> 1R_e$ and $< 1R_e$ within a hexabundle. The top axes are scaled to the effective radius as a fraction of an individual core radius, and effective radius as a fraction of the hexabundles radius. Lower: Distribution of major axis effective radius with redshift for the primary sample. Blue and red lines mark one and half a hexabundle radius respectively. Green points mark the median $R_e$ in bins of 0.01 in redshift. The galaxies with an effective radii of $> 20$ arcsec, were checked by eye, and 9 were not plotted as the effective radii fits were clearly incorrect due to bright nearby stars or galaxies.
}
\label{Re}
\end{figure*}

Figure~\ref{Re} highlights the range of $R_e$ in the survey. We note that no selection is made based on $R_e$ except in the case of extremely large nearby galaxies (see Table~\ref{class}). The median $R_e$ of the primary sample is 4.4 arcsec and 40\% of the galaxies are sampled out to more than $2 R_e$, where we expect any rotation curve to have flattened out. 17\% of the galaxies have $1R_e$ larger than the SAMI bundle, giving higher relative spatial resolution in the centres of these galaxies. The distribution of $R_e$ sampled by the hexabundles does not show a significant trend with redshift.

\begin{figure*}
\centerline{\psfig{file=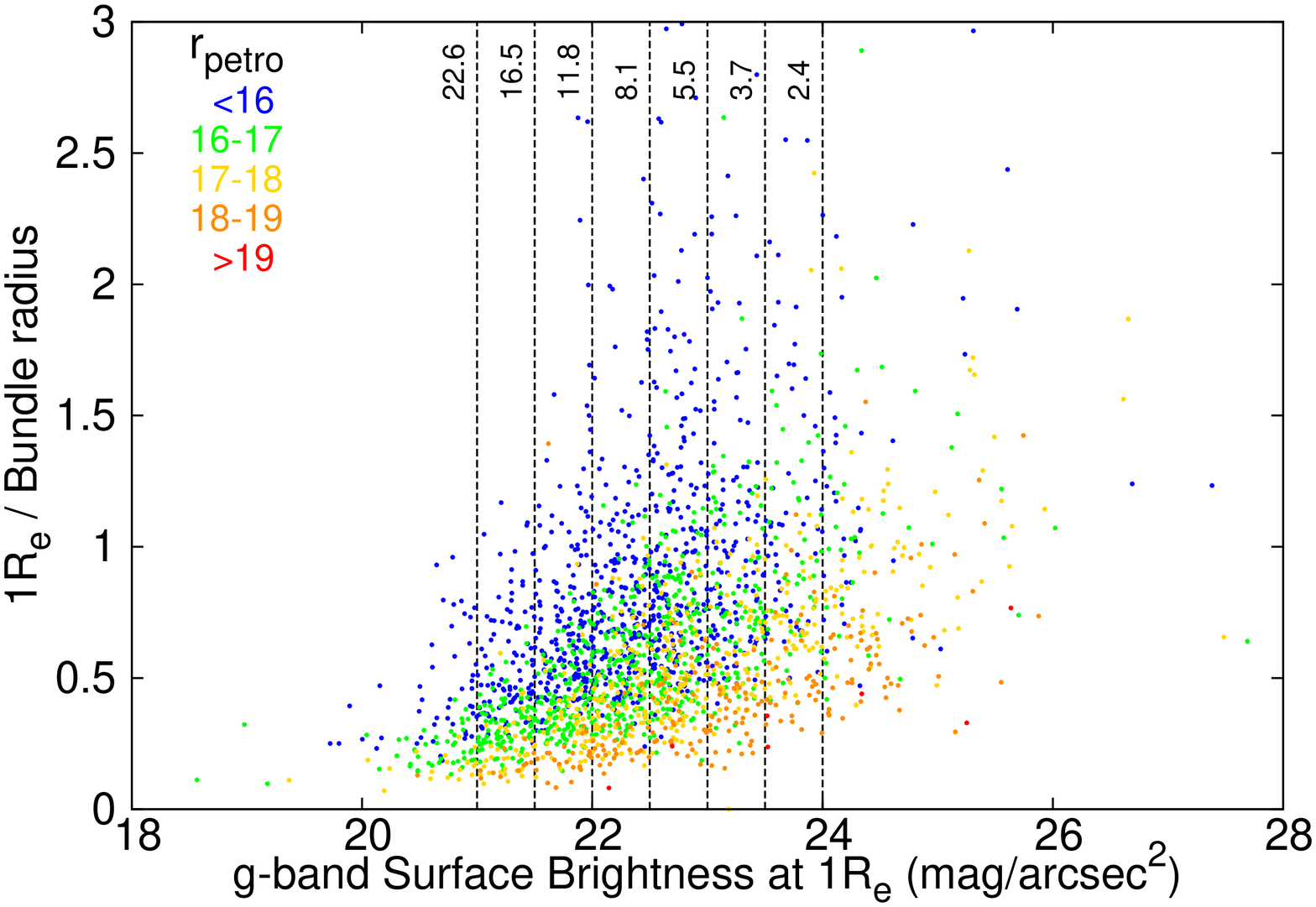, width=8.0cm, angle=-90}}
\caption {Surface brightness in the $g$-band at $1R_{e}$  versus $1R_{e}$ as a fraction of bundle radius. Points are colour coded by the Petrosian $r$ magnitude. Vertical dashed lines list the continuum S/N per \AA\ per hexabundle including fill fraction (not per fibre core, which would be a factor of 0.75 higher), expected to be achieved with 3.5 hours of integration time for the given surface brightness, based on the $B$-band SAMI exposure time calculator. 
The galaxies with lowest surface brightness at $1R_{e}$ tend to be larger, and therefore can be binned to increase S/N.
}
\label{SurfBri}
\end{figure*}

The surface brightness distribution for the GAMA-selected targets is plotted in Figure~\ref{SurfBri}. As expected, the fainter targets have an $R_e$ that is smaller than a hexabundle radius, while brighter targets become more likely to have an $R_e$ larger than the hexabundle, and galaxies of the same $r$ magnitude have fainter surface brightness if the $R_e$ is larger. The surface brightness at $1R_{e}$ is brighter than 23.5~mag\,arcsec$^{-2}$ at $g$-band for 83\% of the sample, which equates to an expected S/N in the continuum of greater than 4 within the 3.5 hour survey integration times (the filling fraction combined with our dithering strategy means each point on the galaxy receives 0.75 times the counts expected from a device with a 100\% filling fraction, which reduces the S/N from the exposure time calculator value of 5). The galaxies with S/N$ < 4$ are primarily  those with faint $r$ magnitudes and small sizes compared to the hexabundle size. For galaxies with low S/N, the spaxels can be binned to increase the S/N in either the continuum or line emission \citep[covariance needs to be accounted for when binning as discussed in detail in][]{All14, Sha14}.

\subsection{Data quality at the extremes of our selection}
\label{quality}

\begin{figure*}
\begin{minipage}[]{0.47\textwidth}
\centerline{\psfig{file=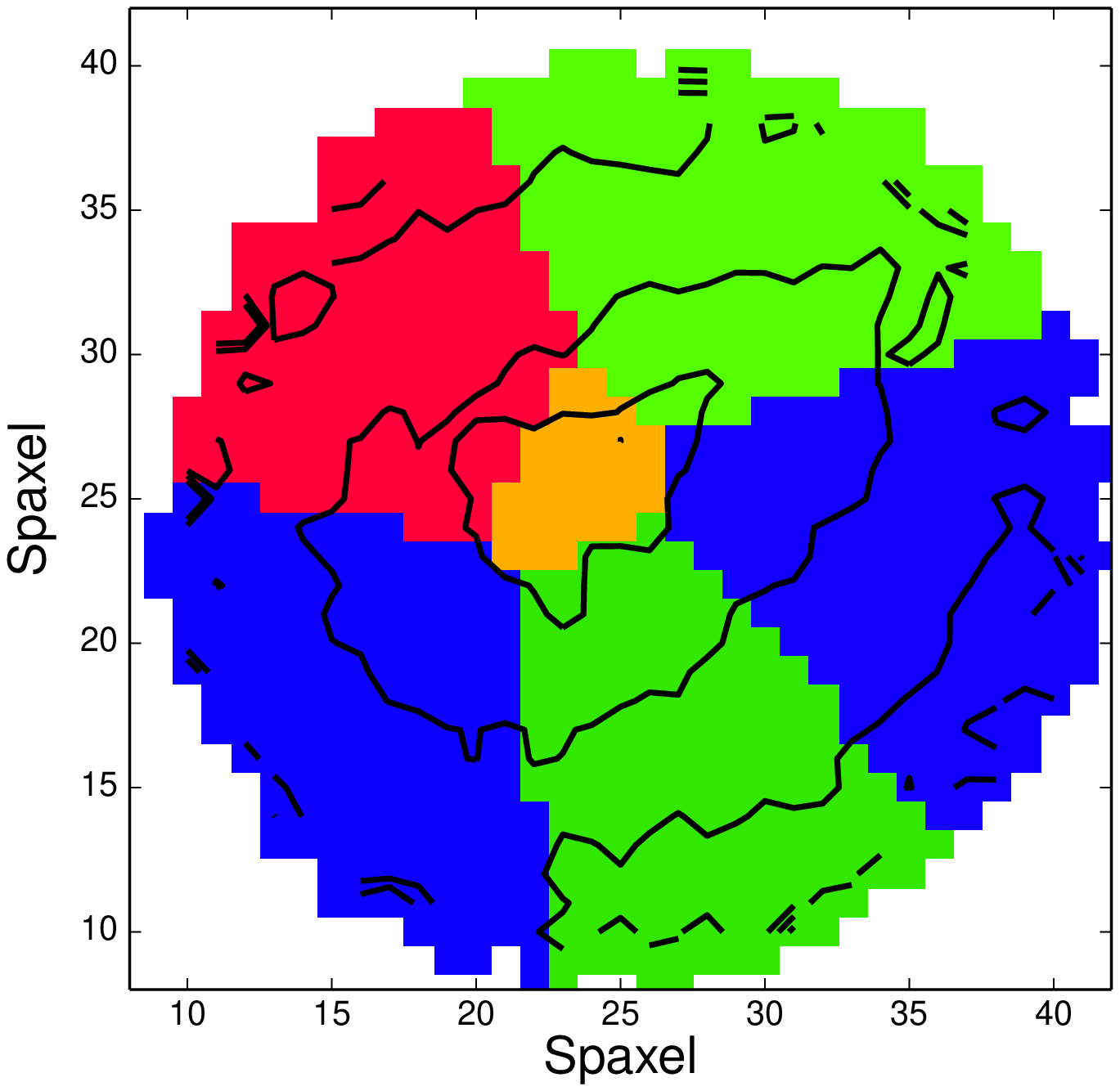, width=10.0cm}} 
\end{minipage}%
\begin{minipage}[]{0.47\textwidth}
\centerline{\psfig{file=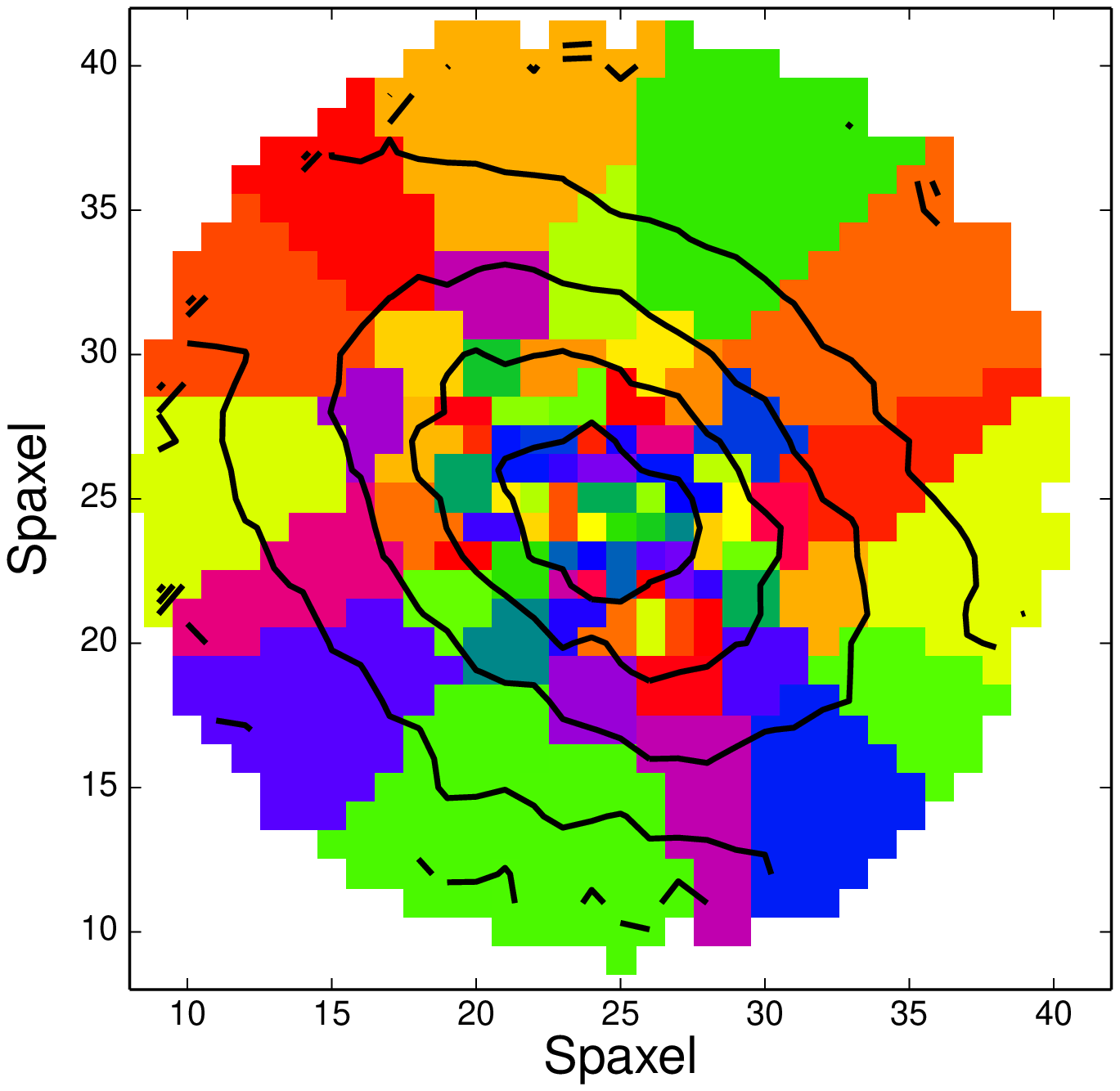, width=10.0cm}} 
\end{minipage}%

\caption{
Examples of the binning required to obtain a minimum continuum signal-to-noise ratio of 5 per \AA\ \citep[accounting for covariance; for details see][]{All14, Sha14}. Each colour region is one bin. We show a low redshift, low mass ($z = 0.015, M_{*} = 10^{8.59} M_{\odot}$) galaxy (left) and a high redshift, high mass ($z = 0.056, M_{*} = 10^{10.15} M_{\odot}$) galaxy (right). SAMI data cubes are re-gridded onto 0.5 arcsec output square spaxels \citep[see][for details]{All14, Sha14}. Continuum flux is shown by the contours.}
\label{SN_cont}
\end{figure*}

\begin{figure*}
\begin{minipage}[]{0.47\textwidth}
\centerline{\psfig{file=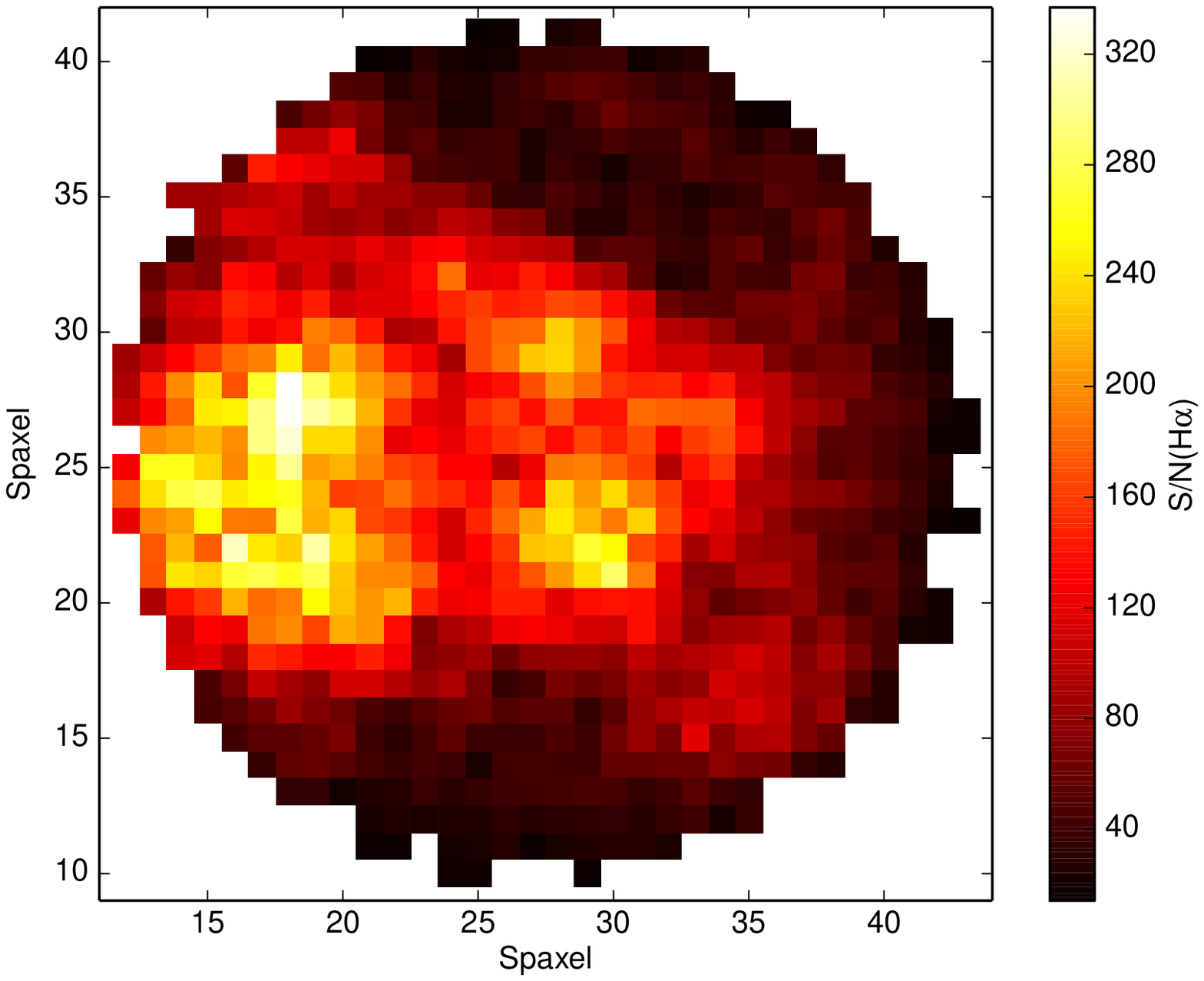, width=9.0cm}} 
\end{minipage}%
\vspace*{3mm}
\begin{minipage}[]{0.47\textwidth}
\centerline{\psfig{file=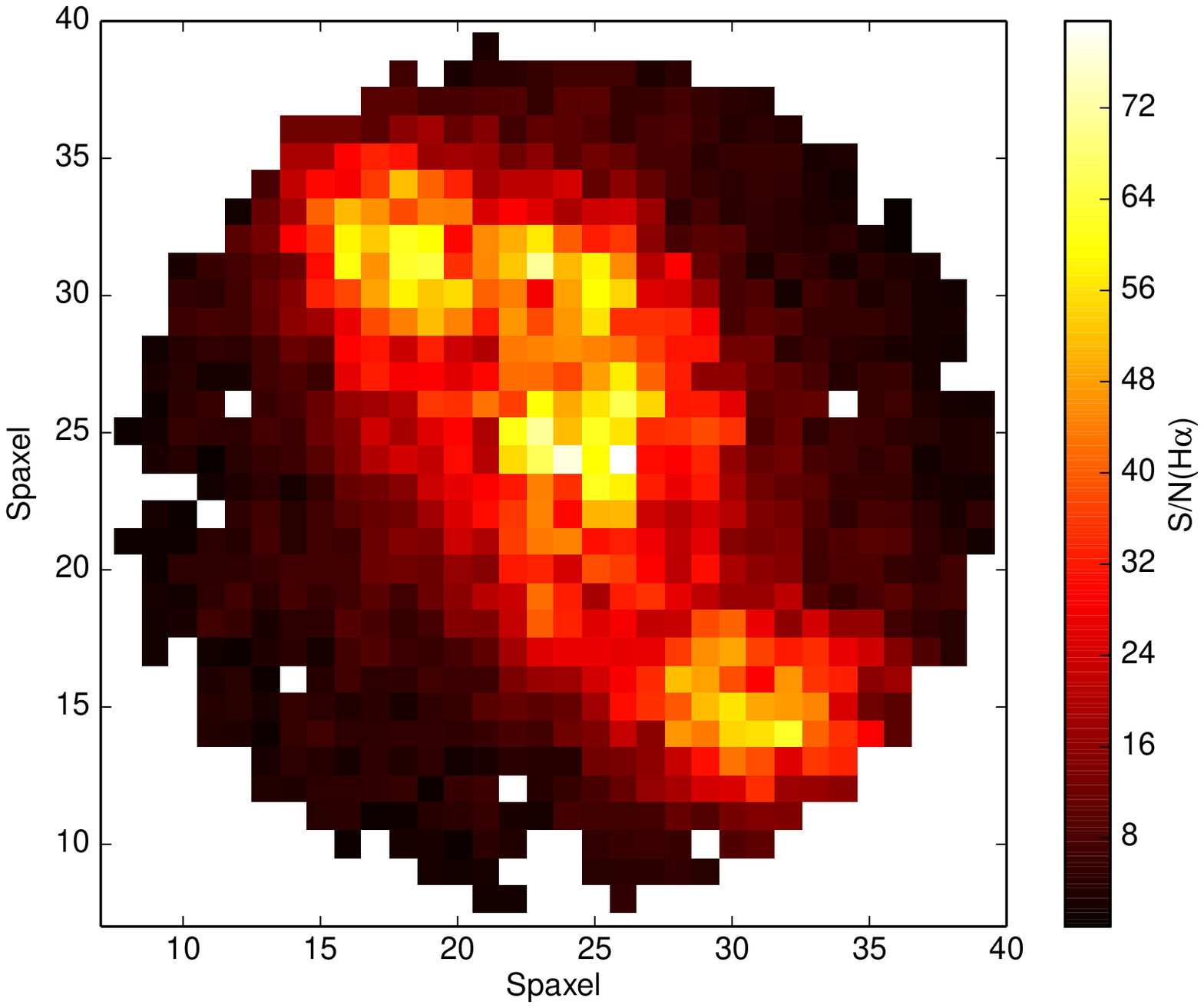, width=9.0cm}} 
\end{minipage}%
\caption{
S/N of H-alpha emission per spaxel for a low z (left; $z = 0.00516$ and $M_{*} = 10^{7.96} M_{\odot}$) and high z (right; $z = 0.08352 $ and  $M_{*} = 10^{11.14} M_{\odot}$) galaxy.}
\label{SN_em}
\end{figure*}

Using early SAMI galaxy survey data from semester 2013A, we have tested that the S/N attainable for both emission and absorption line work at the extremes of our selection, in order to confirm the viability of these limits. Figures~\ref{SN_cont} and~\ref{SN_em} show the signal-to-noise achieved in 
both continuum for stellar fits and in the emission line fits.  The tightest constraints are on stellar kinematics and stellar population fitting, which require S/N in the continuum of $\sim5$ \citep[see, for example][]{Fog2014} and $>10$/\AA\ (confirmed with tests on synthetic spectra) respectively . While these S/N limits are easily achieved in the centres of many SAMI galaxies, for radii greater than half the bundle radius, binning of spaxels is required for stellar population fitting. 
Our adopted redshift range has little impact on the continuum S/N, because the highest stellar mass objects in each redshift bin have similar S/N values out to at least $z\sim0.06$ \citep[see][for a detailed discussion of S/N]{All14}. However the lower stellar mass objects at each redshift have lower S/N as expected, illustrating that lowering the stellar mass cuts further will reduce the fraction of galaxies that can achieve stellar science goals. 

Currently, the lowest stellar mass objects require binning. For example, Figure ~\ref{SN_cont} (left) shows a $z = 0.015$ galaxy with a stellar mass of $10^{8.59}M_{\odot}$, which is at the limit of reasonable S/N values for spatially-resolved stellar kinematics. In this case, the spaxels were adaptively binned to a target S/N $= 5$/\AA\ using the Voronoi tessellation algorithm of \citet{Cap03}. This target has the lowest stellar mass for which we could achieve a S/N $= 5$/\AA\ without binning the entire bundle, and therefore we expect galaxies with stellar masses above $10^{9}M_{\odot}$ to have sufficient S/N for the science requiring stellar kinematics at this redshift. Dwarf galaxies with insufficient S/N in the continuum remain in the survey due to the science cases based on the emission lines. At higher redshift, higher continuum S/N is achievable as the stellar mass increases, as shown in the $z = 0.056$ galaxy in Figure~\ref{SN_cont} (right) with stellar mass of $10^{10.15}M_{\odot}$. The central regions of the bundle have S/N$ > 5$/\AA\ in each spaxel, and minimal binning is required to reach S/N$ = 5$/\AA\, giving resolved kinematics easily, and allowing the possibility of resolving stellar populations or finding stellar population gradients in the target galaxies. 

Science cases requiring emission lines typically need S/N $ \gtrsim 5$. 
Observations in 2013A have shown no trend between galaxy stellar mass and the observed S/N in emission lines, which is dominated instead by the line strength and the observing conditions at the time \citep[see][]{All14}. 
The highest redshift galaxies in general fill less of the hexabundle and have low S/N beyond a half hexabundle radius. However, many of the galaxies in the full SAMI sample at high redshift have an $R_e$ that is larger than half a hexabundle radius, as shown in Figure~\ref{Re}, and will have higher S/N than those observed in 2013A. Beyond our highest redshift selection box (cyan box in Figure~\ref{Mstar}), the $R_e$ of the galaxies decreases further, leading to a decline in spatially-resolved S/N that would have limited the achievable science if we had extended to higher redshift. Figure~\ref{SN_em} shows the H$\alpha$ emission line S/N in two of the galaxies already observed that lie at opposite ends of the selection function. The first is at a low redshift of 0.00516 and $M_{*} = 10^{7.96} M_{\odot}$, while the high redshift example has $z = 0.08352 $ and  $M_{*} = 10^{11.14} M_{\odot}$. In both cases the S/N is $>10$ per spaxel in almost all of the hexabundle. These initial results confirm the viability of our selection limits for achieving emission line science goals.

The redshift and stellar mass range selected for the survey therefore can give sufficient S/N in both the stellar continuum and H$\alpha$ emission line to achieve our science goals with sufficient spatial elements in most cases. 

\subsection{Comparison to previous IFU surveys}
 
 The SAMI Galaxy Survey is unique in comparison to previous IFU surveys because it aims
to observe an order of magnitude more galaxies over a large range in both mass and environment. We use the multiplex of the SAMI instrument to increase the galaxy numbers but at lower spatial sampling than previous surveys such as CALIFA and ATLAS$^{\rm 3D}$ which include finer spatial sampling of fewer nearby targets using a single monolithic IFU. For example, the PPAK IFU used by CALIFA has 331 science fibres across 1 galaxy, while in contrast the 12 SAMI hexabundles have 732 science fibres across 12 galaxies with 61 fibres per galaxy. Therefore SAMI adopts a complementary approach that results in a $15''$ field of view across many more galaxies, while CALIFA has a $74\times64$\,sq. arcsec field and ATLAS$^{\rm 3D}$ has a $33\times41$\,sq. arcsec field, but both include fewer galaxies. In this way SAMI has the dominant statistical power, while CALIFA and ATLAS$^{\rm 3D}$ are the high resolution local benchmarks for detailed comparison to the SAMI galaxies.

CALIFA and ATLAS$^{\rm 3D}$ both focus on more massive nearby galaxies with $0.005<z<0.03$ (median 0.015) and $z<0.01$ respectively, while SAMI samples a broader range in stellar mass up to $z\sim0.095$. The peak of the mass distribution for the CALIFA and SAMI samples is similar, but the SAMI sample includes galaxies down to $10^{7.2}$M$_{\odot}$ , with 21\% of the survey targets having less than $10^{9}$M$_{\odot}$ (see Fig.~\ref{Mstar} and ~\ref{combSM}) compared to $<3$\% in CALIFA. ATLAS$^{\rm 3D}$ is selected to have $>10^{9.7}$M$_{\odot}$ \citep{Cap11}. 
 
CALIFA has a uniquely large coverage for each galaxy with 97\% observed to $2r_{50}$ based on a circularised aperture \citep{Wal14}. Alternatively, both SAMI and CALIFA have $R_{e}$ values measured along the major axis so we can compare surveys using $R_{e}$. CALIFA have 50\% of the galaxies sampled out to $>2R_e$\citep{Wal14}.  The median $R_e$ of SAMI galaxies within a hexabundle is $1.7R_e$  (with a 40\% of the sample imaged to $2R_e$; see Fig.~\ref{Re}), however within the redshift range of CALIFA, SAMI samples galaxies to a median of $1.5R_e$.  By contrast, ATLAS$^{\rm 3D}$ images galaxies to a median of $0.9R_e$ \citep{Ems11}.  
 
On the other hand the surveys sample galaxies on differing spatial scales, which are not set by the IFU elements (lenses or fibres) but by the angular scales limited by firstly, the atmospheric seeing and dither schemes and secondly, by the redshift ranges. SAMI's angular scale elements are expected to average $\sim 2.1$ arcsec due to seeing, while ATLAS$^{\rm 3D}$ has  on-average $1.5$ arcsec seeing \citep{Kra11}  and CALIFA has a median angular  resolution of 2.5\,arcsec \citep{Gar14}. Therefore, based on the redshift ranges of each survey, the spatial resolution for SAMI is $0.17-3.7$\,kpc (median 1.65\,kpc), ATLAS$^{\rm 3D}$  is $<0.3$\,kpc and CALIFA is  $0.26-1.5 $\,kpc (median 1\,kpc; \citet{Wal14}).   Most SAMI galaxies are therefore not sampled as finely as ATLAS$^{\rm 3D}$ and 75\% of SAMI galaxies have the $<2.2$\,kpc sampling of CALIFA. 

In terms of spectral sampling, SAMI's resolution of $R\sim1700$ and $R\sim4500$ in the blue and red remains the highest of these IFU surveys, particularly at red wavelengths. By comparison, CALIFA has $R\sim850$ from $3750-7000$\AA\ and $R\sim1650$ from $3700-4700$\AA\, while ATLAS$^{\rm 3D}$ covers $4800-5380$\AA\ at $R\sim1200$. The high spectral resolution of SAMI is ideal for de-blending complex line profiles \citep[see, for example][]{Fog2014, Ho14} across the broad range of galaxy types. 

SAMI therefore addresses a broad parameter space in environment and stellar mass with fewer spatial elements but over many more galaxies. This is unique compared to these previous IFU surveys, and will overlap with the up-coming MANGA survey (Bundy et al., in prep.).

\section{Final selection: Cluster galaxies}
\label{Clusters}

Key science drivers for the SAMI survey require targets covering a range of environments.  The environment of the GAMA-selected galaxies range from field galaxies to groups, with very few having masses $>10^{14.0}$ M$_{\odot}$ (see Figure~\ref{groups}). To extend the survey to higher mass environments therefore requires the addition of galaxy clusters selected to have virial masses $> 1\times10^{14}$ M$_{\odot}$. 

\subsection{Selection of clusters and cluster galaxies}

The full details of the definition of the cluster sample, including spectroscopy of cluster candidates and selection criteria for cluster members is given in a companion paper (Owers et al., in prep). Here we summarise the key selection criteria to highlight how the cluster sample complements the GAMA-selected sample.

Clusters were chosen with an R.A. range of  22--03hrs so that they are observable in the second half of the year, as the GAMA fields are all observable in the first half of the year. The 8 clusters in the SAMI cluster sample are listed in Table~\ref{clusters}, and are marked in Figure~\ref{aitoff}. 
They were picked to overlap with either the SDSS or 2dFGRS to make use of the existing redshift catalogues for selection of cluster members. We measured additional redshifts using AAOmega fed by the 2dF multi-object fibre-feed for cluster candidates that have $r < 19.4$\,mag to reach 90\% completeness in the cluster fields. 

Imaging data is also important, as existing photometry is used for the cluster member selection, and the SDSS or the VST/ATLAS southern survey fields cover these regions. Stellar masses for cluster galaxies were calculated based on this photometry in the same way as for the GAMA fields. Cluster members were allowed within a radius of $< r_{200}$\footnote{$r_{200} = 0.17\sigma_v(r  <  r_{200})/H(z)$ and is iteratively determined using the velocity dispersion ($\sigma_v$) of the members within
$r_{200}$ \citep[see Owers et al. in prep.][for details]{Car97}} or alternatively we used a limit of $0.5^{\circ}$ when $r_{200} < 0.5^{\circ}$. The stellar mass limit for the selected galaxies was set  to be the same as for the GAMA fields (see Figure~\ref{Mstar}), at the redshift of each cluster.

The cluster catalogue objects were visually inspected as for the GAMA galaxies (see Section~\ref{finalGAMA}).
Within the cluster fields alone, initial visual confirmation has removed $< 4$\% of the galaxies. From  the remaining sources, $\sim$600 randomly-selected galaxies will be observed including the brightest cluster galaxies (BCGs), from 8 clusters as detailed in Table~\ref{clusters} with spatial distribution shown in Figure~\ref{aitoff}.

\begin{table*}
\begin{center}
  \caption{Final clusters selected for the SAMI galaxy survey, with J2000 coordinates, redshift, virial masses within $r < r_{200}$ and the origin of the photometric data used for the galaxy cluster membership selection \citep[for details see][, Owers et al. in prep]{Owe09}.
\label{clusters}}
\begin{tabular}{lrrcll}
\hline 
Cluster name & R.A. &Dec. & $z$ & Virial mass & Photometric \\
  & (deg.)  &(deg.)  &  & ($\times10^{14}\,M_{\odot}$) & data \\	
\hline
EDCC0442 & 6.381 & -33.047  & 0.0494 & 4.5 $\pm$ 0.9 & VST/ATLAS \\
Abell0085 & 10.460 & -9.303   & 0.0556  &  15.4 $\pm$ 1.9 & VST/ATLAS \\
         &  &  &  &  & and SDSS \\
Abell0119  & 14.067 & -1.255  & 0.0442 & 10.1 $\pm$ 1.1 & SDSS \\
Abell0168 & 18.740 & 0.431  & 0.0448 & 3.2 $\pm$ 0.5 & SDSS \\
Abell2399 & 329.389 & -7.794  & 0.0582 & 6.0 $\pm$ 0.8 & SDSS  \\
Abell3880 & 336.977 & -30.575   & 0.0579 &2.8 $\pm$ 0.6 &VST/ATLAS  \\
APMCC0917 & 355.398 & -29.236  & 0.0509 & 2.0 $\pm$ 0.5 & VST/ATLAS \\
Abell4038 & 356.895 & -28.125  & 0.0297 & 2.9 $\pm$ 0.6 & VST/ATLAS \\
\hline 
\end{tabular}
\end{center}
\end{table*}

\section{Combined field and cluster sample properties}
\label{sec_comb}

\subsection{Sky coverage}

\begin{figure*}
 \centerline{\psfig{file=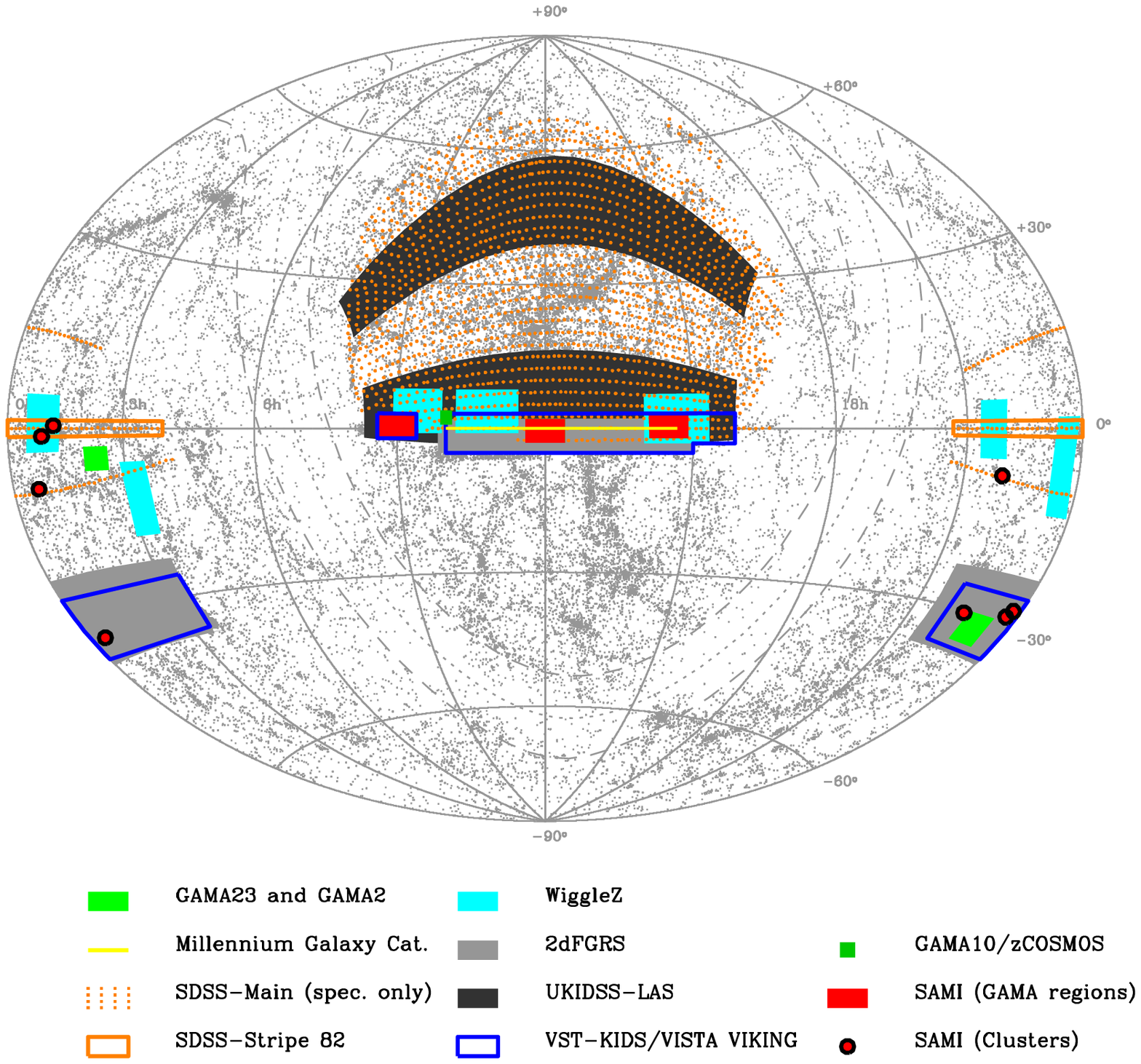, width=14cm}}
   \caption{ Aitoff projection of R.A. (hours) and Dec. (degrees) showing the distribution of the SAMI field and group galaxies in the GAMA regions and the SAMI clusters compared to other related imaging and spectroscopic galaxy surveys including the Millennium Galaxy Catalogue \citep{Lis03}, SDSS \citep{Yor2000, Aba09},  2dFGRS \citep{Col01}, Wigglez \citep{Dri10}, UKIDSS \citep{Law07}, VST KiDS \citep{deJ13} and VISTA VIKING (Driver et al., in prep.). Grey dots mark objects with measured redshifts at $z < 0.1$ from the NASA Extragalactic Database (NED). }
      \label{aitoff}
\end{figure*}

The SAMI Galaxy Survey regions were selected to have supporting spectroscopic data for redshift selection and imaging data, as the photometry is necessary as a proxy for stellar mass in our selection criteria. Figure~\ref{aitoff} shows the sky distribution of the SAMI field and cluster regions compared with other spectroscopic and imaging surveys. The SAMI field galaxy regions are the equatorial fields from GAMA, which also partially overlap several other surveys as detailed in Section~\ref{ancil}. The target selection was based on spectroscopic redshifts and photometry from GAMA in the equatorial regions, while the cluster galaxies overlap either SDSS or VST regions that provide the photometry, and SDSS or 2dFGRS for the redshifts. 

The final catalogue has only one object in common with the ATLAS-3D survey and none in common with the CALIFA mother catalogue.

\subsection{Group and cluster masses}
\label{Environs}

\begin{figure}
\centerline{\psfig{file=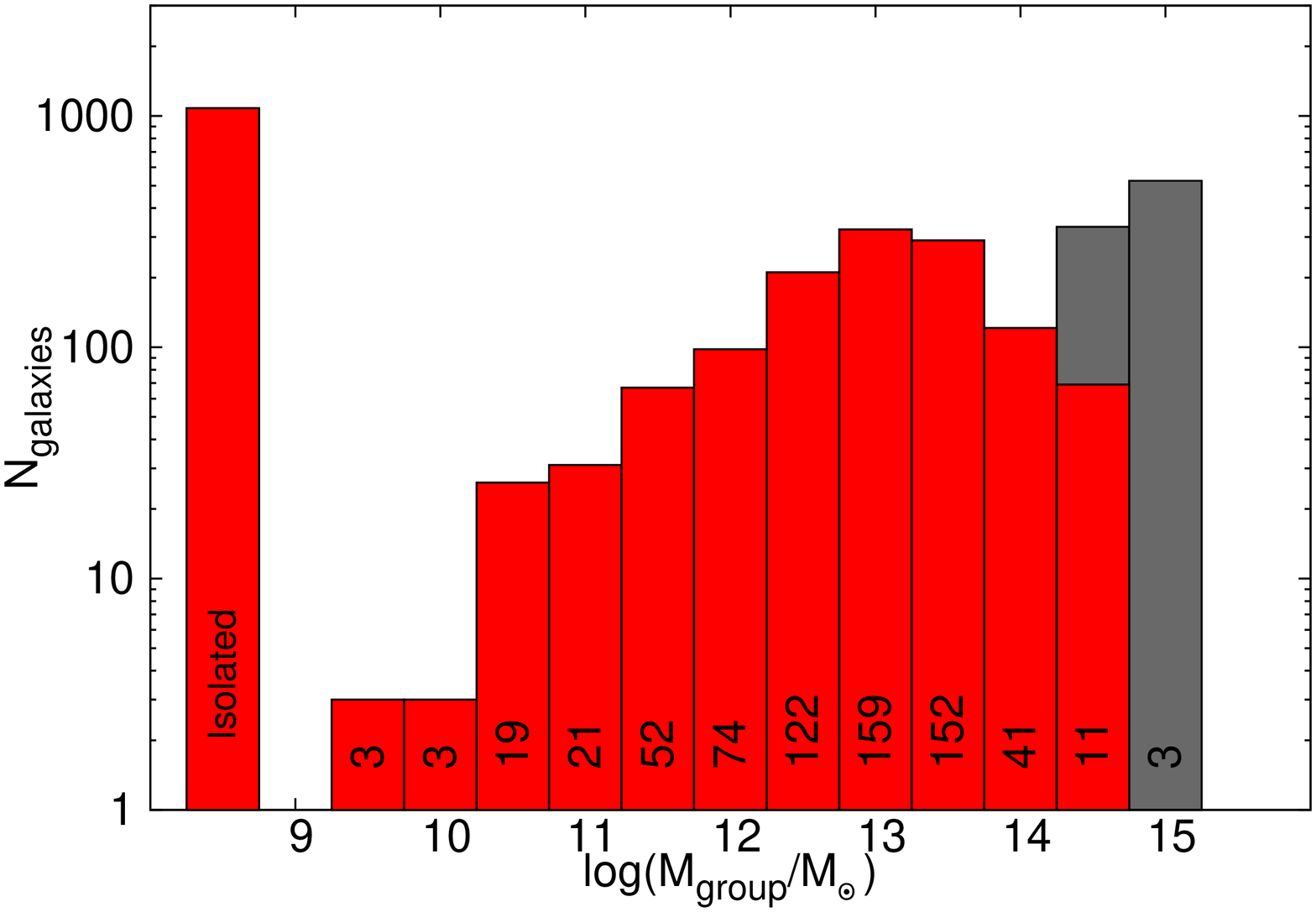, width=6cm, angle=-90}}
   \caption{Histogram of the number of galaxies in groups/clusters of a given dynamical mass for groups/clusters hosting SAMI galaxies. The red histogram shows the SAMI GAMA-region primary samples (corrected to the cosmology in Section~\ref{intro}). Cluster virial masses are shown in the grey histogram for the galaxies in 8 clusters. In the bin centred on 10$^{14.5}$ M$_{\odot}$ where the environment masses overlap, 69 galaxies (red) are from the SAMI GAMA-region sample and 278 (grey) are from 5 of the clusters. Galaxies in the remaining 3 clusters  lie in the highest mass bin. The survey will observe 90\% of the total SAMI GAMA-region targets and 600 of the cluster galaxies. The black numbers list the number of groups or clusters in each bin. In some cases only one galaxy in a GAMA group is in the SAMI catalogue, in which case the number of galaxies equals the number of groups in that case.}
      \label{groups}
\end{figure}

We take advantage of the GAMA group catalogue (\citealp{robotham11}, which includes $\sim$97\% of galaxies 
in the SAMI GAMA-region selection catalogue) to characterise the typical environments covered by our sample. 
In Figure~\ref{groups}, we show the distribution of group masses for all SAMI targets. Galaxies not assigned 
to a group are indicated as isolated systems. SAMI GAMA-region targets cover almost the entire range of environments 
found in the local Universe (apart from rich clusters), but appears particularly well suited to study isolated systems 
and galaxy groups with masses in the range 10$^{12.5} < $M$_{group}/M_{\odot} < 10^{13.5}$  due to the selection function of groups in the GAMA sample. 
Interestingly, this range of group masses is where the cold gas content and star formation activity 
of galaxies starts to be affected by the environment \citep{catinella13}, making SAMI an ideal dataset to 
study how nurture influences the star formation cycle of galaxies. 
As the GAMA survey regions do not include many clusters of galaxies at low redshifts,  the 
SAMI GAMA-region sample predominantly contains galaxies residing in groups with masses of
less than $\sim$10$^{14}$ M$_{\odot}$. However, the addition of the SAMI cluster sample extends the mass range, making the full SAMI survey 
a unique dataset to study galaxy evolution across {\it all} environments. Although rich clusters only contain a very small fraction of the total stellar mass \citep[$< 2$\% for $>10^{14.5} h^{-1}{\rm M}_{\odot} $ halos, see][]{Eke05}, it is important to include them to establish which physical processes are unique to the rich cluster environment and which are ubiquitous.

 \subsection{Galaxy stellar masses}
 
\begin{figure}
\centerline{\psfig{file=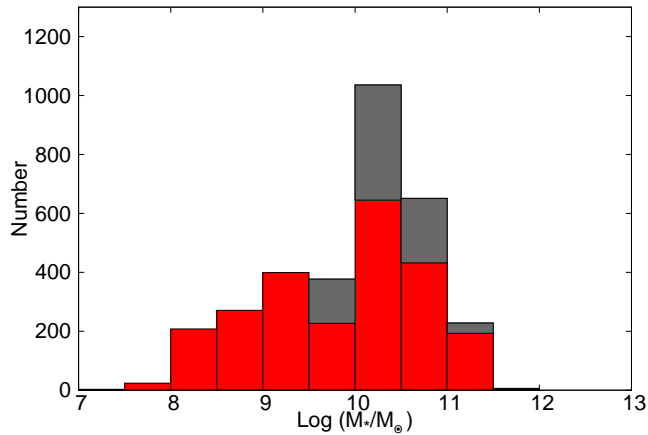, width=6cm, angle=-90}}
 \caption{Combined stellar mass distribution of the SAMI survey primary sample (red). The component of that distribution from the clusters is in grey, while the remainder are from the GAMA regions. The completed survey aims to observe 90\% of these GAMA region targets and 600 of the cluster galaxies.}
   \label{combSM}
\end{figure}

 Figure~\ref{combSM} shows the combined stellar mass distribution of the primary targets in both the GAMA and cluster samples. Filler targets extend to lower stellar masses and are not plotted. The stellar mass cut-offs for the cluster targets were set by the same limits as the GAMA regions and hence there are no cluster galaxies selected below 10$^{9.5}$ M$_{\odot}$. The full survey covers a broad range in stellar mass primarily from 10$^{8}$ to 10$^{11.5}$ M$_{\odot}$, from dwarfs to massive BCGs. The sample allows for a direct comparison of the high-density environment cluster galaxies to galaxies in low-density environments in the field.
 
 \subsection{Colours and magnitudes}

\begin{figure}
\centerline{\psfig{file=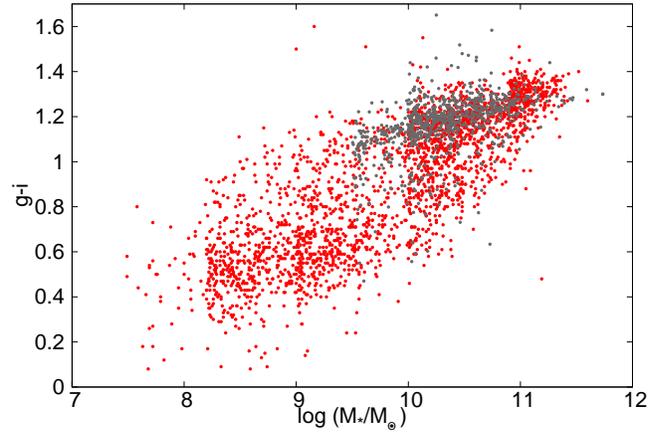, width=6cm, angle=-90}}
\caption{Stellar mass versus $g-i$ colour for the SAMI primary sample in the GAMA regions  (red), and four of the clusters, for which aperture-matched photometry is available at the time of writing  
(grey; see Owers et al., in prep., for details). 
The magnitudes are the $mag$\_$auto$ values from the GAMA catalogue, which come from sextractor fits.
}
\label{col-mag}
\end{figure}

The colour-mass plot in Figure~\ref{col-mag} illustrates how the SAMI GAMA-region targets span from the blue cloud, across the green, to the red sequence. This broad distribution will be crucial for SAMI survey studies of gas and stellar evolution in galaxies and the impact of environment on evolution from the blue to the red sequence. Galaxy morphologies change with colour, and the SAMI survey is clearly sampling a wide range of galaxy types. The cluster galaxies, on the other hand, are primarily on the red sequence, as expected from the stripping of gas in the cluster environment, which quenches star formation, reddening the galaxy.

\section{Ancillary data from other surveys}
\label{ancil}

\subsection{UV, optical and infrared}
\label{ancilOpt}

The SAMI Survey primary sample is selected from the GAMA survey, which by design has extensive
multi-wavelength coverage from the FUV to the far-IR (see Driver et al., in prep.). The base $ugriz$ data drawn from SDSS have been reprocessed and astrometrically
aligned with the Visible and Infrared Survey Telescope for Astronomy
(VISTA) Kilo-Degree Infrared Galaxy (VIKING) survey data (Driver et al., in prep), enabling matched
aperture photometry in $ugriZzYJHK$ \citep[i.e. identical apertures,
deblending solutions etc;][]{Hil11}. The GAMA regions used by SAMI have
also been extensively observed with GALEX, as part of the Medium
Imaging Survey \citep{Mar05a} and augmented through dedicated
observing campaigns led by the GAMA team (\url{http://www.mpi-hd.mpg.de/galex-gama/}). Stacked drizzled WISE data
has been prepared by the WISE team \citep[procedures described in][]{Jar12,Clu14}, achieving greater depth and resolution than that provided in
the WISE all-sky data release. Finally the Herschel-Atlas survey
\citep{Eal10}, an extensive Herschel Space Observatory
survey, also targeted the GAMA regions used by SAMI, providing
observations with the Photodetector Array Camera and Spectrometer (PACS) and the Spectral and Photometric Imaging Receiver (SPIRE) completing the multi-band
(multi-facility) coverage comprising of 21 broad-band filters. The
$5\sigma$ limiting AB mag depths of these filters are: $FUV = 24.5$;
$NUV = 24.0$; $u = 22.1$; $g = 23.0$; $r = 22.7$; $i = 22.7$; $z = 20.8$; $Z = 23.1$; $Y = 22.4$;
$J = 22.1$; $H = 21.2$; $K_s = 21.3$; W1 = 21.1; W2 = 20.4; W3 = 18.6; W4 = 16.6;
100$\mu$m = 13.0; 160$\mu$m = 13.4; 250$\mu$m = 12.0; 350$\mu$m = 12.2;
450$\mu$m = 12.5 AB mags (for full details see Driver et al., in prep). 

These data allow for complete broad-band spectral analysis providing
both robust stellar mass measurements \citep{Tay11} as well as
dust mass, dust temperature \citep[e.g. via MAGPHYS in][or similar]{daC08}, and global star-formation rate estimates
independent of aperture corrections. At the present time further
observations are underway with ESO's VST imaging facility as part of
the Kilo-Degree Survey \citep[VST KiDS;][]{deJ13} which will
significantly improve the depth (by $\sim 2$ mag) and spatial resolution
(0.7$''$) in the $ugri$ bands.

The spectroscopic component of the GAMA survey consists of $\sim180,000$
redshifts \citep[and spectra, see][]{Hop13} within the SAMI 
regions enabling the construction of a robust halo catalogue \citep{robotham11} extending down to $10^{11}$M$_{\odot}$ (based on velocity
dispersion measurements and calibrated against numerical simulations),
along with a variety of environmental markers based on nearest
neighbour distances, local density measurements \citep{Bro13},
and information as to whether the selected SAMI galaxy resides in a
void, filament or tendril \citep{Alp14a,Alp14b}. The high
spectroscopic completeness of the GAMA survey \citep[$\sim98$\%; ][]{Dri11} ensures that these environmental markers are robust \citep[see halo mass confirmations by][]{Alp12, Han14}. Additional
analysis derived from either the Public GAMA Data Release 2
(Liske et al. in prep.), or directly from the full GAMA database, also
include surface profile analysis with GALFIT3 \citep[see][]{Kel12}
providing either 9-band single S\'ersic fits or where appropriate,
bulge-disc or bulge-bar-disc decompositions. Finally as the SAMI Main
Survey sample is embedded within GAMA, the rarity or normality of each
SAMI galaxy is known a priori. In effect GAMA can be used to place the
results from the SAMI Main Survey sample into a cosmological context.

\subsection{Radio data}
\subsubsection{Continuum}

\begin{figure}
\centerline{\psfig{file=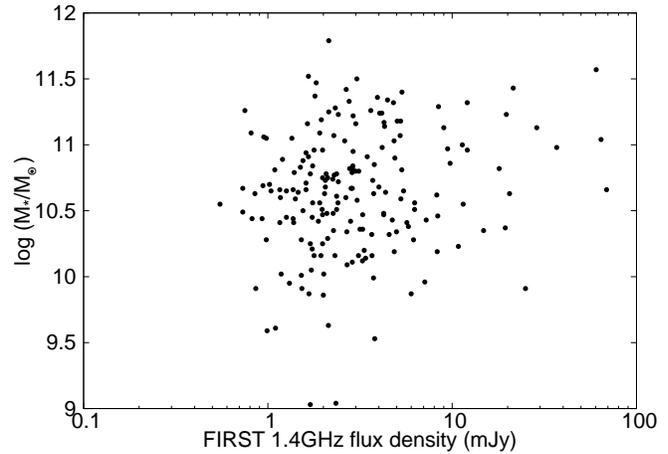, width=6.0cm, angle=-90}}
\vspace*{3mm}
\caption{Stellar mass versus 1.4GHz flux density for the sources from the SAMI GAMA-region catalogue, that have reliable detections in FIRST.
}
\label{FIRST}
\end{figure}

The GAMA data includes radio observations from the GMRT at 325MHz \citep{Mau13}. These regions have also been covered by the 1.4 GHz Faint Images
of the Radio Sky at Twenty-cm \citep[FIRST;][]{Bec95} and NRAO
VLA Sky Survey \citep[NVSS;][]{Con98} catalogues. The FIRST catalogue has a 5 arcsec resolution, but NVSS's larger beam (50 arcsec) can lead to confusion with nearby sources.
There are 216 galaxies in the SAMI Galaxy survey (GAMA regions) with FIRST or NVSS detections that have been confirmed as associated sources. The catalogues primarily detect the higher stellar mass objects and therefore only provide data for SAMI galaxies above $10^{9.5}$ M$_{\odot}$ with a median of  $10^{10.65}$ M$_{\odot}$ as illustrated in 
Figure~\ref{FIRST}. Within the 2 dex in stellar mass, we cover 2 dex in 1.4\,GHz flux density, and future observations are planned to extend the radio detections to galaxies of lower stellar mass. The spatially-resolved SFR can be used to test the physics underpinning the radio-FIR correlation in star-forming galaxies as a function of both stellar mass and environment. 

The SAMI cluster regions are covered by FIRST/NVSS in the equatorial regions and the Sydney University Molonglo Sky Survey \citep[SUMSS;][]{Mau03} and NVSS for the four southern clusters.

\subsubsection{Radio 21cm H{\sc i}}

\begin{figure}
\centerline{\psfig{file=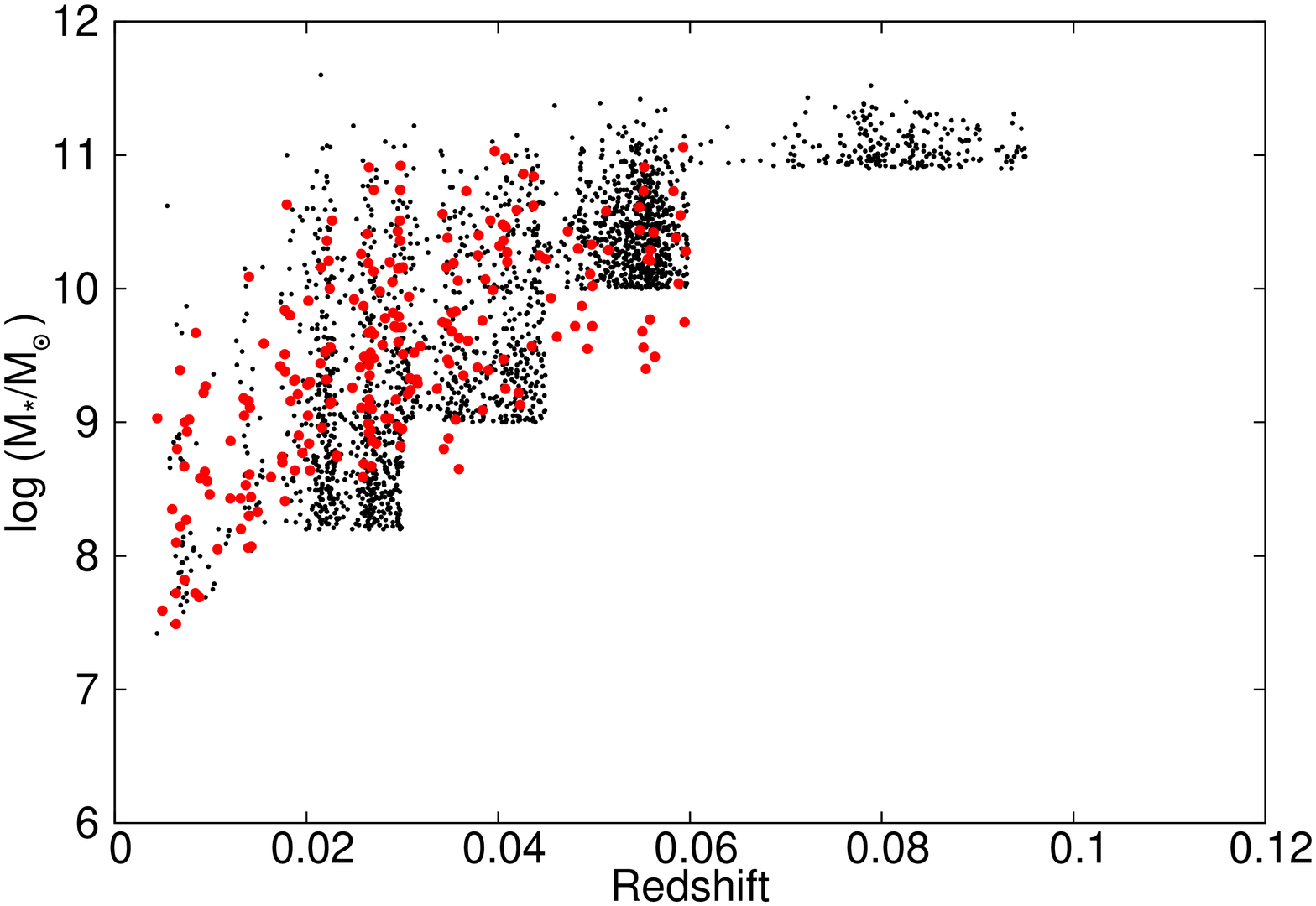, width=6.0cm, angle=-90}}
\centerline{\psfig{file=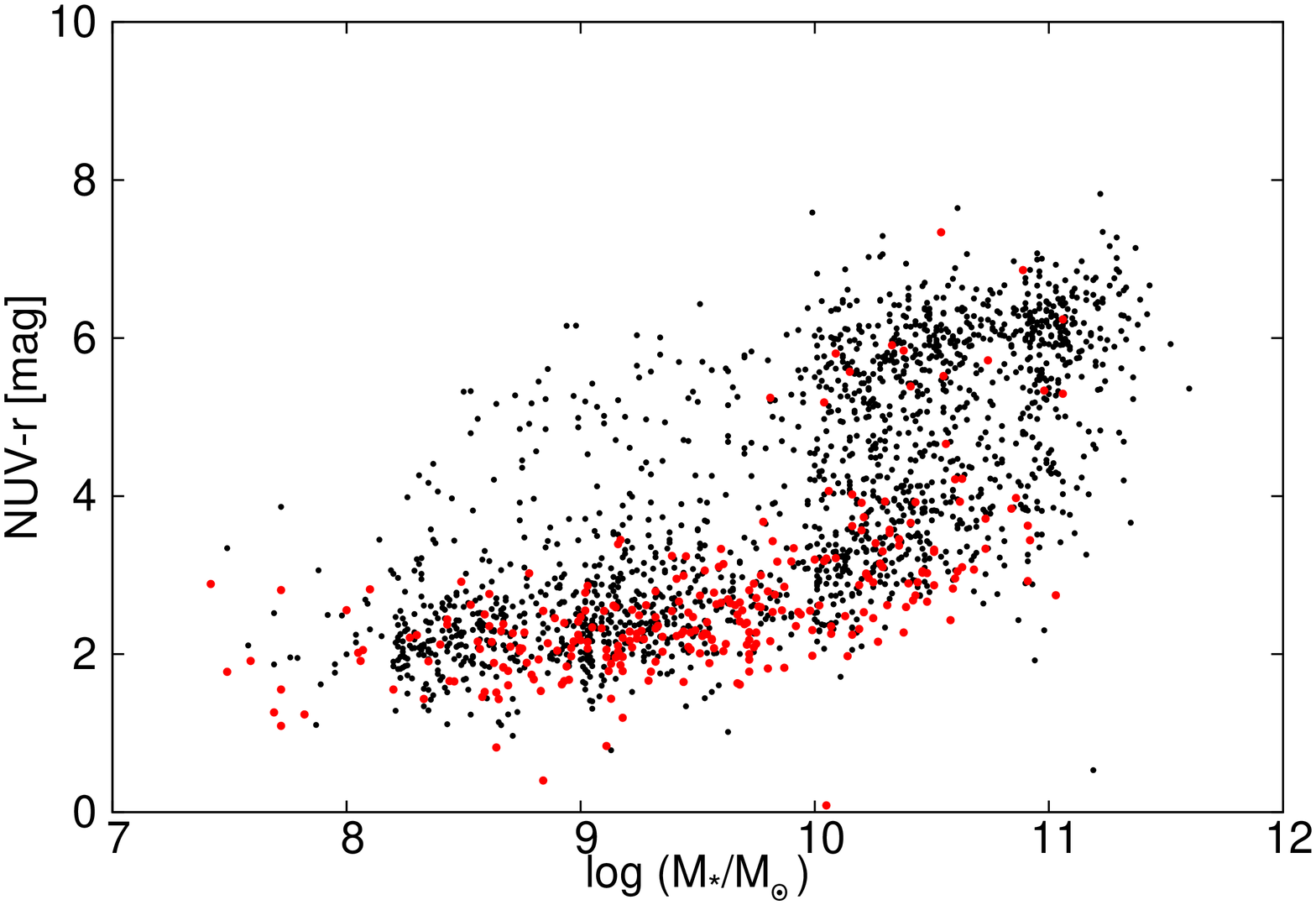, width=6.0cm, angle=-90}}
\centerline{\psfig{file=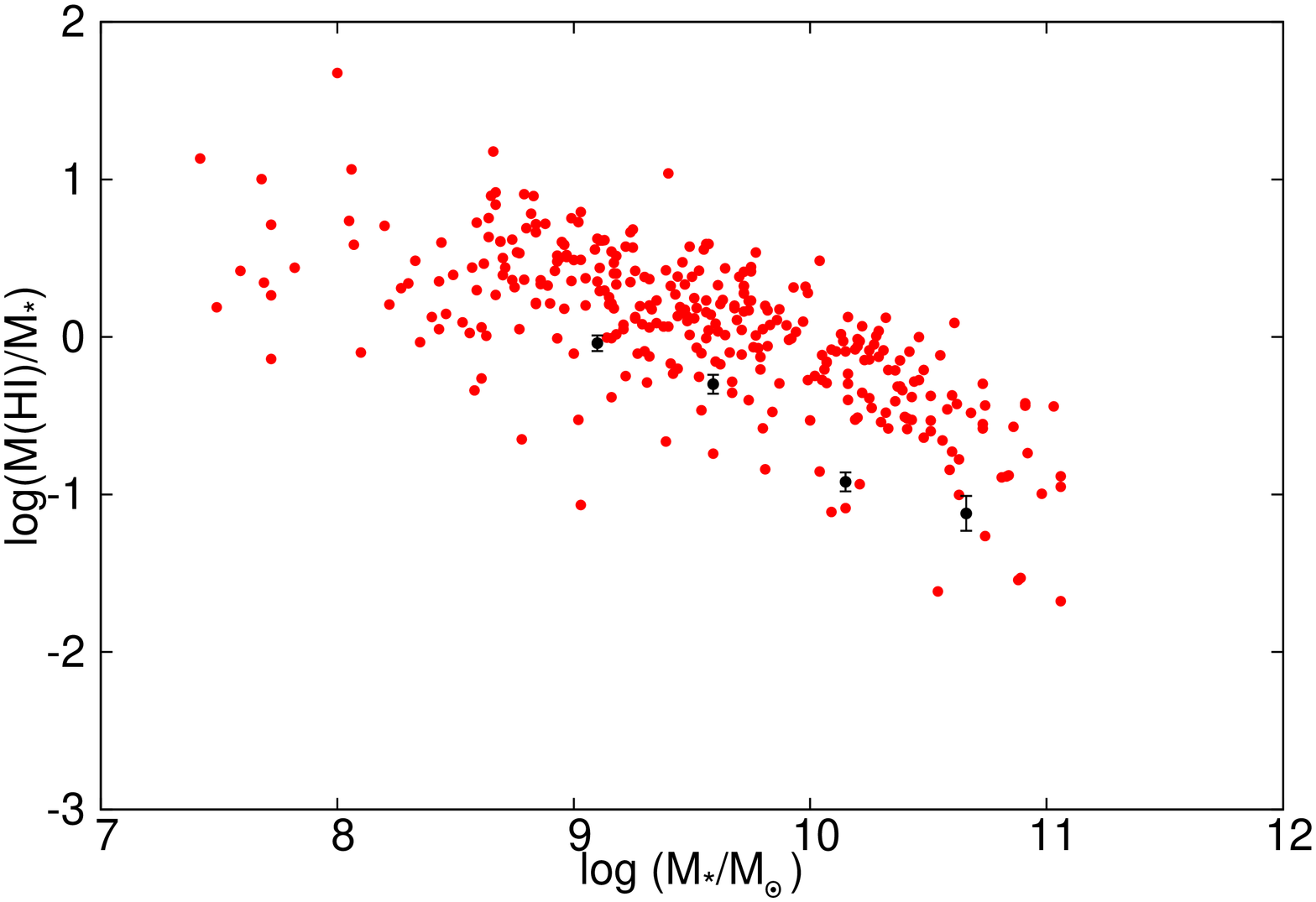, width=6.0cm, angle=-90}}
\vspace*{3mm}
\caption{Distribution of the SAMI survey targets that have observations from the ALFALFA survey, in stellar mass vs redshift (top) and
$NUV-r$ colour versus stellar mass (middle) for the SAMI GAMA-region primary sample (black dots). Galaxies detected 
   by ALFALFA are indicated by the red dots. Some SAMI filler targets have ALFALFA detections and lie outside the primary target range. Lower: The $M(H${\sc i}$)/M_{*}$ vs. stellar mass relation for SAMI galaxies 
   detected by ALFALFA. The average scaling relations for \hi-normal galaxies in the local Universe \citep{cortese11}
   are indicated by the black dots with error bars.
}
\label{ALFALFA}
\end{figure}

Soon the Australian
Square Kilometre Array Pathfinder \citep[ASKAP; ][]{Joh07} will survey 21 cm emission from H{\sc i} in the equatorial GAMA fields. In the meantime, 21\,cm \hi\ line observations for part of the SAMI sample in the GAMA regions are already available thanks 
to the Arecibo Legacy Fast ALFA (ALFALFA) survey \citep{alfaalfa05}. ALFALFA is a state-of-the-art 
blind \hi\ survey, covering the high galactic latitude extragalactic sky visible from Arecibo ($\sim7000$ square degrees) up to 
$z\sim$0.06. ALFALFA observed all the Arecibo Spring night sky above Declination +0, thus including $\sim$58\% 
of the GAMA survey regions. The typical rms noise limit of ALFALFA is $\sim$2.4 mJy/beam at a velocity 
resolution of 10\,km s$^{-1}$ \citep{haynes2011}, corresponding to a 6.5$\sigma$ limit \citep{saintonge07} in \hi\ gas 
mass of $\sim$1$\times$10$^{10}$ M$_{\odot}$ at $z = $0.05, for a velocity width of 200\,km s$^{-1}$. 
However, at such low declination the gain of the 
Arecibo telescope is lower than nominal, implying a slightly higher ($\sim$20\%) rms. 
A preliminary list of H{\sc i} ALFALFA sources for the GAMA region (M. Haynes priv. comm.) has been cross-matched with the SAMI galaxies using a 15 arcsec aperture between the ALFALFA identified optical counterpart and the GAMA positions, as well as a redshift difference between the optical and the H{\sc i} counterpart less than 0.001. Figure~\ref{ALFALFA} (top) shows where the resulting 249 matched galaxies sit in the full sample. It is notable that the full range of stellar masses is covered by the ALFALFA matches.

Figure~\ref{ALFALFA} (middle) shows the position of SAMI galaxies detected by ALFALFA in a $NUV-r$ colour versus stellar mass diagram. 
\hi\ detected galaxies are among the bluest galaxies in our sample. This is simply a consequence 
of the fact that, at the average redshift of SAMI, ALFALFA detects only the most \hi-rich objects. 
Indeed, compared with the average \hi\ scaling relations of local galaxies \citep{cortese11}, the SAMI galaxies 
in the ALFALFA catalogue clearly occupy the gas-rich envelope of the $M(H${\sc i}$)/M_{*}$ vs. stellar mass 
scaling relation of nearby galaxies (see Figure~\ref{ALFALFA}, lower).

\section{SAMI Galaxy Survey Observing program}
\label{Observing}

The SAMI Galaxy Survey has been awarded long-term status on the AAT. 
Observations began in semester 2013A and will continue until
the end of 16A, totalling 151-181 nights (depending on scheduling) of dark time.
The aim is to complete 90\% of the primary GAMA-region targets (2164 galaxies), and 600 cluster galaxies. Filler targets will bring the total observed to $\sim3400$ galaxies.

The SAMI instrument has 13 hexabundles, so each field observes 12 galaxies plus one standard star. Two sky fibres per hexabundle are mounted in separate connectors to sample the sky at 26 positions across the $1^{\circ}$ diameter field of view of the AAT prime focus. A 42m fibre cable joins the 793 hexabundle fibres and 26 sky fibres  to the AAOmega spectrograph, located in the coud\'{e}  room at the base of the telescope. AAOmega is used with the dichroic at 570nm,
the 580V blue arm grating with a central/blaze wavelength of 4800A and the 1000R 
red arm grating with a central/blaze wavelength of 6850A.
 
\begin{figure}
\centerline{\psfig{file=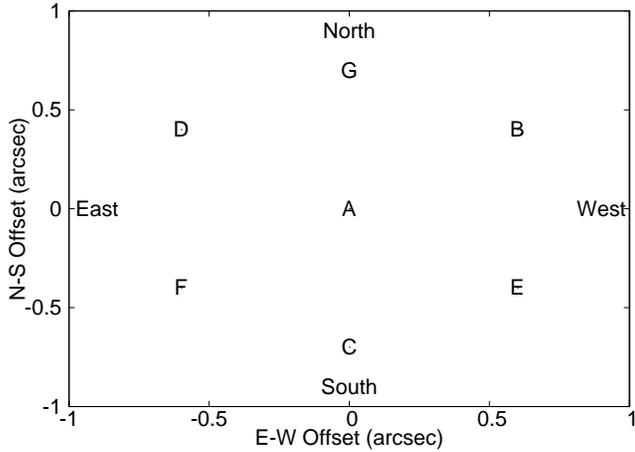, width=6.0cm, angle=-90}}
\vspace*{3mm}
\caption{Distribution of the 7-point dither positions (A-G) on-sky. All fields are observed in each of these positions.
}
\label{dither}
\end{figure}

 Integration times are set to 7 frames of 1800s each, giving 3.5 hours on-source. Each frame is dithered by between 0.4 to 0.7 arcsec in a set pattern as shown in Figure~\ref{dither} \citep[see][for details of the optimisation of the dither scheme]{Sha14}. The pattern includes a central position, a north and a south offset and 4 radial positions. The exposure times were chosen based on the SAMI S/N calculator\footnote{\url{https://www.aao.gov.au/science/instruments/sami/}} 
with an aim to achieve a continuum S/N $> 10/$\,\AA\  to a surface brightness of 22.6 and 22.1 in the Vega $B$ and $R$-bands respectively.  Each galaxy plate has the holes drilled for 2 fields, and the hexabundles are replugged into the holes for the new field at the end of each field observing. Typically 2 fields can be completed in a night.

\subsection{Flux calibrators and guide stars}

\subsubsection{Spectrophotometric standard stars}
Primary spectrophotometric standards are observed twice per night by aligning individual hexabundles on to each of two standard stars (selected from the ESO standards list) in turn.

One secondary photometric standard is observed by one of the hexabundles for each plate (for each 12 galaxies). The simultaneous observation allows for accurate spectrophotometry irrespective of observing conditions. However, the on-sky density of SDSS photometric standard stars is insufficient for the effective configuration of SAMI fields. Therefore, these spectrophotometric standards are chosen from SDSS imaging to be similar in colour to an F-star (to give a spectrum which is smooth near the telluric features), based on the equation 

\begin{multline}
([(u-g) - 0.82]^2 + [(g-r) - 0.30]^2 + \\ [(r-i) - 0.09]^2 + [(i-z) - 0.02]^2)^{0.5} < X
\label{spec_colour1}
\end{multline}
for the GAMA and cluster fields.

The priority of the stars for tiling purposes is set by the colour value, X, which is given in Table~\ref{Priority_stars} along with the magnitude
cut-off based on the extinction-corrected $r$-band PSF magnitude. Then a range of SDSS flags are checked, to ensure the flags for the stars are consistent with the object:

\begin{itemize}
\item having zero velocity ({\sc stationary} in {\sc object2} {\it objc} flag)
\item being a primary observation in the SDSS full survey ({\sc survey\_primary} in {\sc resolve\_status} flags)
\item having a photometric observation ({\sc photometric} in {\sc calib\_status} flags)
\item not having contamination from other sources, saturated pixels or poor sky subtraction (not {\sc blended, too\_many\_peaks, cr, satur, badsky} in {\sc object1} {\it objc\_flags})
\item having a position away from saturated pixels, unchecked regions or other spurious features (not\,{\sc peaks\_too\_close,\,notcheched\_centre, satur\_centre, interp\_centre, psf\_flux\_interp}  in {\sc object2} {\it objc\_flags2}).
\end{itemize}

\begin{table}
\begin{center}
\caption{Priorities for selection of standard stars based on the psf $r$-band magnitude and the colour values X, defined in Equation~\ref{spec_colour1}. The higher priority value stars will be tiled first, and if none are available to match a field, then lower priorities are accepted by the tiling algorithm.
\label{Priority_stars}}
\begin{tabular}{lll}
\hline 
Priority & $r_{psf}$& X \\
\hline
8 & $\leq 17.25 $& $ < 0.08$\\
7 & $\leq 17.25$ &  $0.08\leq X < 0.16$\\
5 & $17.25 < r_{psf}\leq17.5$ &  $< 0.16$\\
4 & $17.5 < r_{psf}\leq17.75$ & $< 0.16$\\
3 & $\leq 17.25$ & $0.16\leq X < 0.2$ \\
2 & $17.25 < r_{psf}\leq17.5$ & $< 0.2$ \\
1 & $17.5 < r_{psf}\leq17.75$ & $< 0.2$ \\
\hline 
\end{tabular}
\end{center}
\end{table}

\subsubsection{Guiding and guide stars}

\begin{figure}
\centerline{\psfig{file=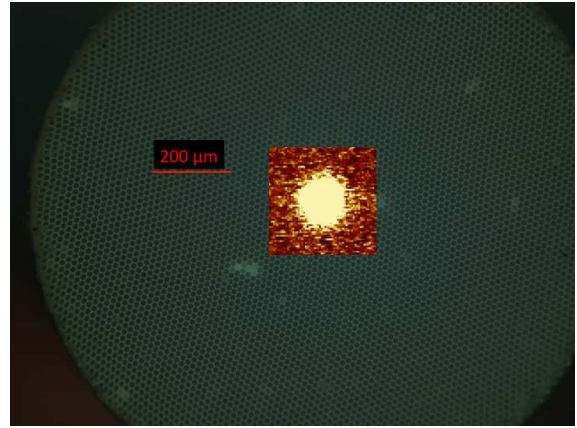, width=7.5cm}}
\caption{Microscope image of the front face of a polymer guide bundle with an inset image of a star through the guide bundle, at the same scale. The guide bundle field of view (22.8 arcsec) is large compared to the stars, making acquisition simple.
}
\label{GBpic}
\end{figure}

At the start of the SAMI Survey, guide stars were limited to be at the centre of each field, because
the old guide camera was mounted to see through a central hole in the field plate. However, the SAMI instrument was further upgraded in mid-2013 to use new coherent polymer fibre guide bundles. The polymer bundles, shown in Figure~\ref{GBpic}, are made from a flexible multi-core polymer, with over 7000 cores each of 16$\mu$m diameter, and an outer diameter of 1.5\,mm including cladding (see Richards et al, in prep.). They are mounted in magnetic connectors similar to those of the hexabundles. Three guide bundles can now be positioned anywhere in the $1^{\circ}$ diameter field, simplifying the choice of guide stars for each plate. 

The main advantage of guiding with the bundles over the previous guide camera is that the guide camera was mounted on a gantry above the field plate, and there was some flexure in that gantry that affected the guiding position when the telescope was at large zenith distances. The new guide bundles are mounted directly in holes in the plate and have no flexure. Furthermore the polymer bundles have 70-80\% transmission, nearly doubling the throughput, and allowing fainter guide stars, which in turn are easier to allocate to each field. The guide bundle field of view is 22.8 arcsec, which means minimal precision is required to ensure the star is visible in the bundle.
Three guide stars are selected for each field because three polymer bundles are imaged by the guide camera simultaneously. The telescope guides off the star that is closest to the centre of the distribution of the galaxies in the $1^{\circ}$ diameter field.

Guide stars were selected in the $g$-band where the guide camera sensitivity peaks. In order to be bright enough to guide on through the polymer bundles, the brightness was limited to $9 < g < 14.5$, with colours of $-0.5 < (g-r)$ or $(r-i) < 2.0$ and UCAC4 \citep{Zac13}
proper motions below 15 mas yr$^{-1}$.

\subsection{Optimal tiling of the SAMI fields, and plate production}
\label{sec_tiling}

\subsubsection{{\it Greedy} tiling}

The aim is to observe the SAMI Galaxy Survey primary catalogue targets to a completeness of 90\%. To do so requires efficient configuration of objects per observing plate, and this is done using the {\it Greedy} algorithm \citep{Rob10}.  The {\it Greedy} algorithm is an heuristic algorithm designed to tile densely-packed surveys in a way that gives the highest completion rate. \citet{Rob10} presents two tiling options, {\it Dengreedy} and {\it Greedy}, and we have adopted {\it Greedy} tiling. The former places tiles based on the position that has the lowest survey completeness, while the latter maximise the total number of objects within a field of view. If there is a large dynamic range of sky density (as in SAMI, where between 0 to 4 SAMI pointings are required in each position), {\it Greedy} performs better. However when the dynamic range is small (as in GAMA) {\it Dengreedy} is preferable.

All of the objects in both the GAMA regions and cluster regions have an assigned priority in the survey catalogues (PRI\_SURV, see Table~\ref{catdata}). Higher priority targets are tiled first, and then lower priorities are used to fill the 12 hexabundles per plate. 

Each SAMI plate requires 12 galaxies from the SAMI galaxy catalogue, 1 secondary standard star and 3 guide stars within a $1^{\circ}$ diameter field of view. These targets need to be selected such that the holes to be drilled in the plate have sufficient separation to prevent the hexabundle magnetic connectors from touching, requiring 15mm of spacing, equivalent to 228 arcsec. 

The tiling chooses the best location to place a field based on the target density within a SAMI field of view. In some cases, as the survey progresses, there will not be 12 targets available in each pointing, and therefore the number of field pointings on sky is not simply the target number divided by 12, but is dependent on the source distribution and most efficient tiling method. 

\begin{figure*}
\centerline{\psfig{file=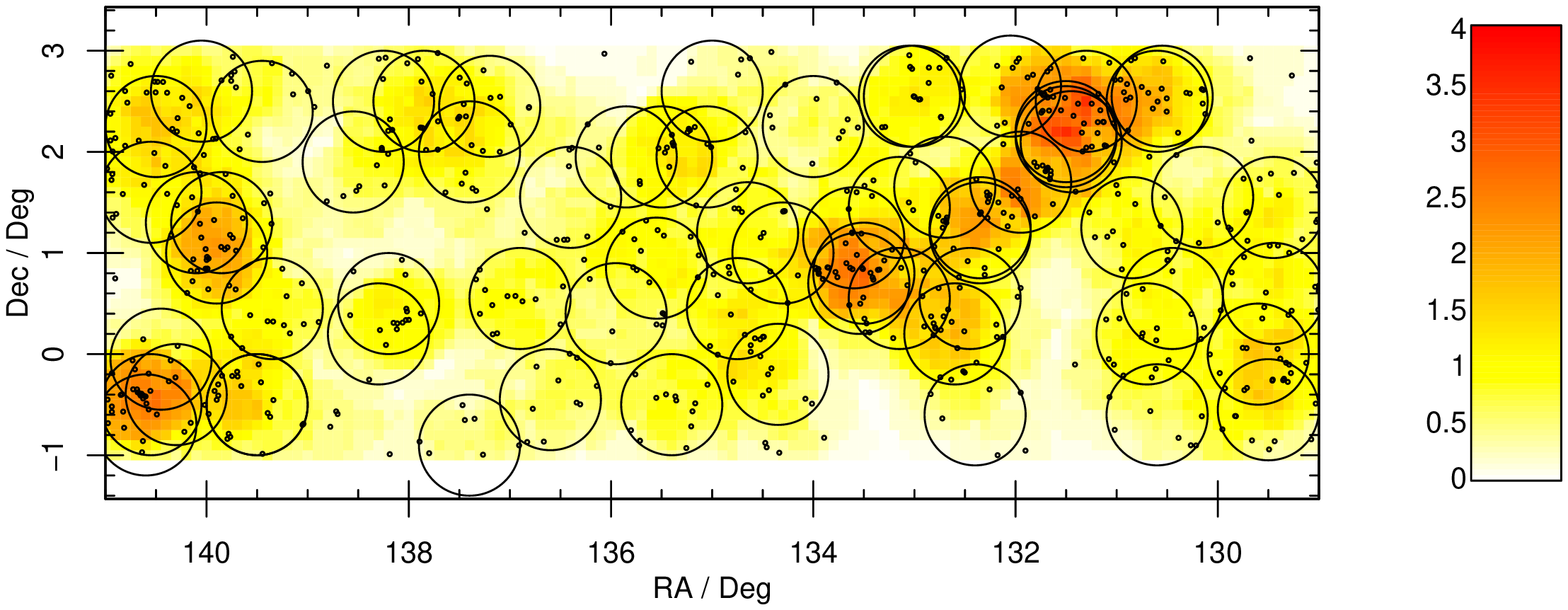, width=6.0cm, angle=-90}}
\vspace*{3mm}
\centerline{\psfig{file=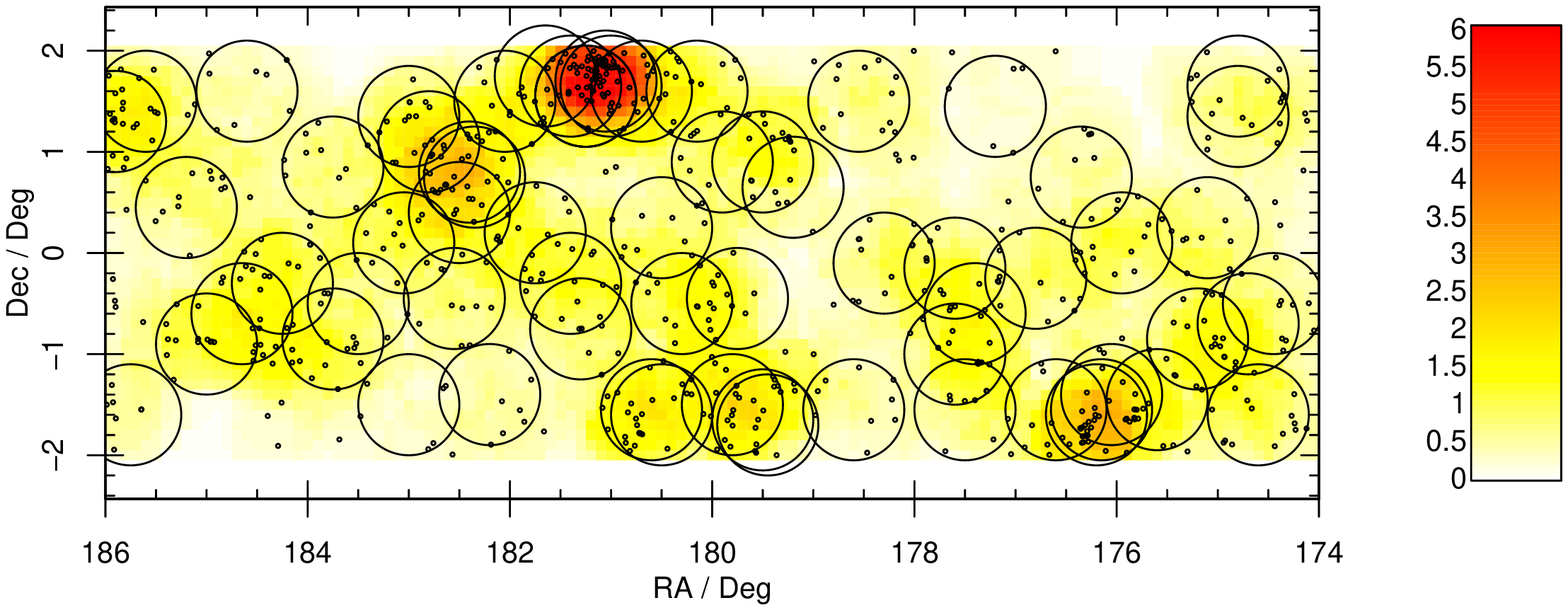, width=6.0cm, angle=-90}}
\vspace*{3mm}
\centerline{\psfig{file=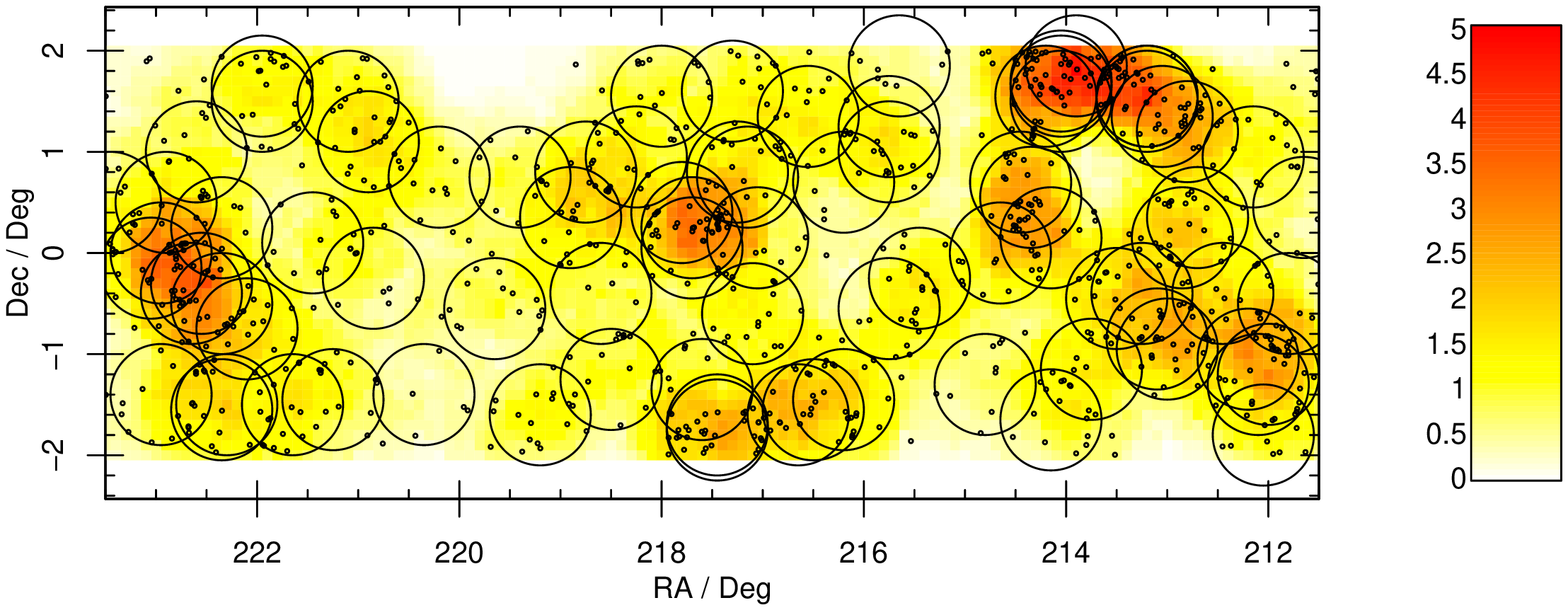, width=6.0cm, angle=-90}}
\caption{Simulation of tiling the SAMI survey in the three GAMA regions G09 (top), G12 (middle) and G15 (bottom). The catalogue sources are shown as small black circles. The colour bar shows the number of SAMI pointings required to observe all objects at that location in R.A./Dec. For example, if there are 48 main survey objects within a $1^{\circ}$ field of view, as in the highly-clustered regions, then the corresponding bin value is 4 (colour-coded red). In order to most efficiently reach 90\% completeness (based on the {\it Greedy} algorithm), tiles need to be placed at the position of each large black circle, representing the $1^{\circ}$ diameter SAMI field.}
\label{fig_tiling}
\end{figure*}

Using the {\it Greedy} algorithm, we simulated the tiling of the full survey in each of the three GAMA regions, and in the clusters. The goal is to observe 90\% of the primary GAMA target list, which is reached after 67, 67 and 84 tiles in the G09, G12 and G15 regions respectively as shown in Fig~\ref{fig_tiling}. 
This simulation shows the resulting observations will then include $> 2170$ primary 
and $> 260$ filler galaxies in the 218 fields. 
Allowing for repeat observations of $\sim 100$ objects (for quality control or repeats of data taken in poor conditions), we plan to observe additional fields, up to 237 pointings giving 2800 galaxies  in total. A further 58 fields are required to observe 600 primary cluster targets with 60 lower priority galaxies and 30 repeats.

In order to confirm that the tiling process does not introduce a bias with respect to local density, we used the 5th nearest neighbour surface densities from the GAMA database \citep{Bro13}, to show that the percentage of galaxies that are isolated (taken to be surface density $<1$) and the fraction in groups (taken as surface density $>10$) were in agreement (within statistical errors) between the objects tiled in the simulation compared to those in the parent sample.

\subsubsection{Configuration of the Plug Plate}

SAMI's 13 hexabundles, 26 sky fibres and 3 imaging guide bundles are held in the
optical plane by a 24~cm diameter, 3~mm thick, pre-drilled steel plug plate.
Because the plate holes required for one observation take up a small fraction ($< 2\%$) 
of the total plate area, it is possible for multiple fields to be drilled onto a
single plate, thus reducing the number of plates which must be manufactured and
the length of time taken to reconfigure the instrument between the successive
pointings of one night's observation.
A significant part of the process of plate configuration therefore revolves
around ensuring that the targets for each field are chosen such that the holes
for the multiple stacked fields do not overlap, and include sufficient space
between them to accommodate the fibre connectors and the action of plugging and
unplugging them.

The process of configuring a SAMI plate consists of 5 stages:

1) The 2 to 3 survey regions to be observed on a given night are determined,
which defines a pool of targets available for each observing field on the
plate. In addition to the 12 science targets, 1 calibration star, 3 guide
stars, a  field must also have 26 sky positions. Each survey region is divided into several tiles,
which are ordered by the number of high priority targets they contain, with the
first 12 targets being unique to that tile.

2) For each tile, a list of valid candidate fields is produced by overlaying a
grid on the tile area and testing each cell position for suitability as the
plate centre. At each position all possible combinations of the available targets are produced, and those deemed to be valid are retained as candidate fields.
As well as containing the correct number of targets, the targets in a valid field
must meet the proximity constraint of a 228~arcsec (15~mm) minimum separation
between any two targets.

3) The lists of candidate fields for each sky pointing are ordered based on their
mean science target priority. Starting from the field with highest priority,
pairs of valid plate configurations are formed (one field from each of the two
regions to be observed). Similar to the previous stage, if any of the targets
are in conflict (i.e. the targets violate the proximity constraint of
3.8~arcmin) the combination is rejected.

From the resulting candidates the field (or several if multiple plates of the same
regions are required) with the highest mean plate priority is used.

4) Sky fibre position determination: To best utilise the space on the plug
plate, and to reduce the number of fibres which need to be repositioned between
observations, each of the 26 sky positions is chosen such that the position on
the sky for each of the two stacked fields corresponds to a single hole on the
plate. Additionally, to provide optimal sky subtraction data, the positions are
chosen to provide uniform coverage over the available plate area.

This procedure is carried out using the Cone of Darkness software \citep{Lor2014} which
divides the plate area using a grid of cell size 1, $\frac{1}{2}$,
$\frac{1}{3}$, $\frac{1}{5}$ or $\frac{1}{8}$ times the diameter of the plate
holes, depending on the object density of the individual sky region. From the
resulting pool of unoccupied grid cells the one which is most isolated (furthest
from the nearest occupied cell) is tested for suitability as a sky position, by
means of a TAP ADQL \citep{Ort11} 10~arcsec radial proximity search of the
SuperCOSMOS Sky Survey catalogue \citep{Ham2001}. If the catalogue yields no
objects within the search radius, the candidate position is allocated to a sky
fibre. Otherwise the candidate is discarded. In both cases the corresponding
grid cell is removed from the candidate pool, the next most isolated cell is
identified and the process iterates until suitable positions for all 26~sky
fibres are found.

\begin{figure*}
\centering
\includegraphics[width=12cm]{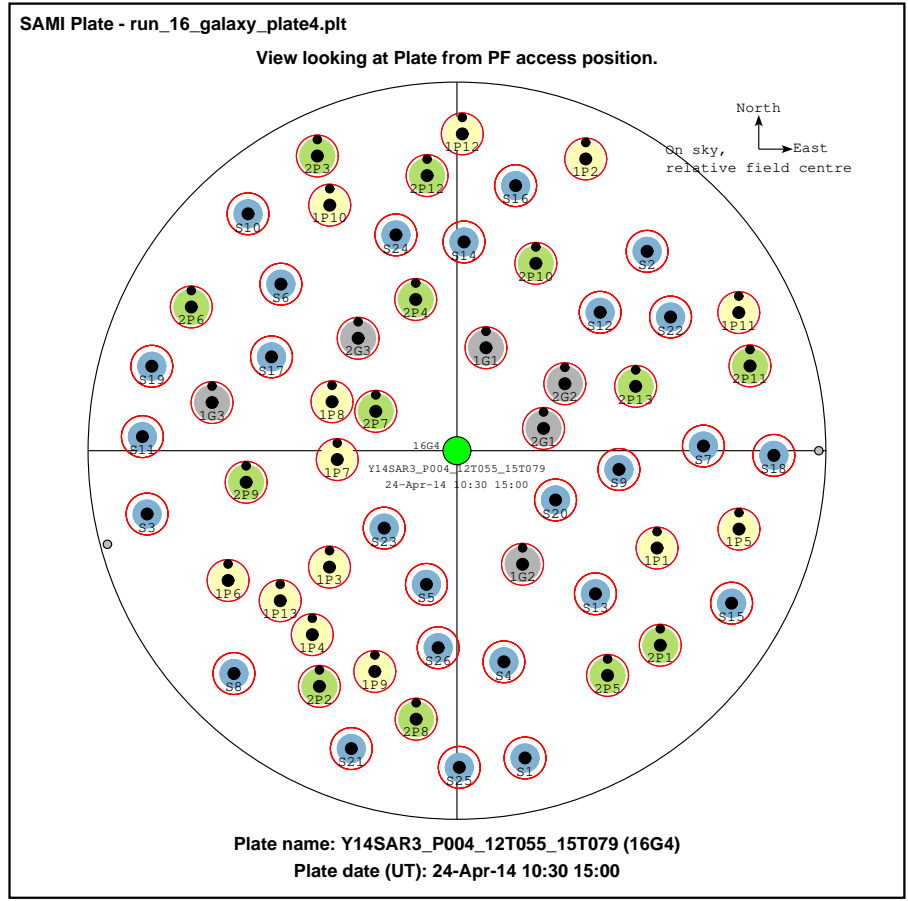}
\caption {Schematic of a typical SAMI plug plate showing the positions and
relative sizes of the holes (black). The science targets are depicted in yellow
(field 1) and green (field 2) with their corresponding guide stars in grey. The
26 sky positions common to both fields are shown in blue. The radius of the
filled circles corresponds to the physical size of the fibre connectors. The red
circles represent the exclusion area around each position. The central black hole in each red circle is where the sky fibre or hexabundle sits. The second smaller black dot to the north of each science and guide bundle hole is a separate smaller hole which accommodates the key which sets the rotation of the hexabundles.}
\label {figPlateConfiguration}
\end{figure*}

5) Once all the plate fibre locations have been defined, the AAT prime focus
astrometric model is applied to the positions, correcting for optical distortion
of the telescope and instrument and atmospheric differential diffraction based
on expected meteorological conditions at the time of observation, observing
wavelength and the representative hour angle of the observation. A final
correction is made to account for the difference in the estimated temperature of
the plug plate at fabrication ($23^{\circ}{\rm C}$ in Summer and 
$16^{\circ}{\rm C}$ in Winter) and during the observations ($15^{\circ}{\rm C}$ 
in Summer and $10^{\circ}{\rm C}$ in Winter).  A schematic of a typical SAMI
plate, showing the positions of the target, guide and sky holes is shown in
Figure~\ref{figPlateConfiguration}.

\section{Summary}
\label{summary}

The SAMI Galaxy Survey began observations at the AAT in 2013 and will map spectral gas and starlight across 3400 galaxies within 3 years. Motivated by the aim to cover a broad range in stellar mass and environment at $z < 0.095$, targets for the survey were primarily selected from the GAMA survey, including field and group galaxies, and supplemented by 8 galaxy clusters.  The three equatorial GAMA regions cover 144 square degrees in total. After balancing the tradeoffs between absolute magnitude and stellar-mass selection, the sample was selected using a well-defined and robust proxy for stellar mass. The primary sample from the GAMA regions consists of 4 volume-limited samples, a low mass sample, a higher-redshift filler sample, and several lower mass filler samples.  Clusters were chosen in regions with SDSS or VST/ATLAS photometry, and lie between $0.03 < z < 0.06$. Here we have presented the characteristics of the SAMI galaxy sample, the observing method for the SAMI Galaxy Survey, and coverage in other wavebands. We have also shown the instrument throughput and improvement due to the SAMI-II upgrade. Both the instrument and target selection have been crafted to maximise the science gains from our ambitious galaxy survey, which is the largest IFU galaxy sample to date.

\subsection*{Acknowledgements}

The SAMI Galaxy Survey is based on observations made at the Anglo-Australian Telescope. The Sydney-AAO Multi-object Integral-field spectrograph (SAMI) was developed jointly by the University of Sydney and the Australian Astronomical Observatory. The SAMI input catalogue is based on data taken from the Sloan Digital Sky Survey, the GAMA Survey and the VST ATLAS Survey. The SAMI Galaxy Survey is funded by the Australian Research Council Centre of Excellence for All-sky Astrophysics (CAASTRO), through project number CE110001020, and other participating institutions. The SAMI Galaxy Survey website is http://sami-survey.org/ .

We would like to thank the Australia Astronomical Observatory and University of Sydney instrumentation groups for their support and dedication to making the SAMI instrument.
The SAMI survey has greatly benefitted from the excellent technical support offered by the AAO in Sydney and by site staff at the Anglo-Australian Telescope.

GAMA is a joint European-Australasian project based around a spectroscopic campaign using the Anglo-Australian Telescope. The GAMA input catalogue is based on data taken from the Sloan Digital Sky Survey and the UKIRT Infrared Deep Sky Survey. Complementary imaging of the GAMA regions is being obtained by a number of independent survey programs including GALEX MIS, VST KiDS, VISTA VIKING, WISE, Herschel-ATLAS, GMRT and ASKAP providing UV to radio coverage. GAMA is funded by the STFC (UK), the ARC (Australia), the AAO, and the participating institutions. The GAMA website is: http://www.gama-survey.org/.

SMC acknowledges the support of an Australian
Research Council Future Fellowship (FT100100457). CJW acknowledges support through the Marie Curie Career Integration Grant 303912. LC acknowledges support under the Australian Research Council's Discovery 
Projects funding scheme (DP130100664). MSO acknowledges the
funding support from the Australian Research Council through a Super
Science Fellowship (ARC FS110200023). ISK is the recipient of a John Stocker Postdoctoral Fellowship from the Science and Industry Endowment Fund (Australia)

We thank Riccardo Giovanelli, Martha Haynes and the ALFALFA
team for access to available ALFALFA data in advance of publication.

Based on data products (VST/ATLAS) from observations made with ESO Telescopes at the La Silla Paranal Observatory under programme ID 177.A-?3011(A,B,C).

This research has made use of the NASA/IPAC Extragalactic
Database (NED), which is operated by the Jet
Propulsion Laboratory, California Institute of Technology,
under contract with the National Aeronautics and Space
Administration. 

Funding for SDSS-III has been provided by the Alfred P. Sloan Foundation, the Participating Institutions, the National Science Foundation, and the U.S. Department of Energy Office of Science. The SDSS-III web site is http://www.sdss3.org/.

\vspace{3mm}
\hrule
\vspace{3mm}
\noindent
$^{1}$ Sydney Institute for Astronomy (SIfA), School of Physics, The University of Sydney, NSW 2006, Australia \\
$^{2}$ Australian Astronomical Observatory, PO Box 915, North Ryde, NSW 1670, Australia;\\
$^{3}$ ARC Centre of Excellence for All-sky Astrophysics (CAASTRO); \\
$^{4}$ ICRAR, The University of Western Australia, Crawley WA 6009, Australia; \\
$^{5}$ School of Mathematics and Physics, University of Queensland, Brisbane, QLD 4072, Australia;\\
$^{6}$ SUPA, School of Physics and Astronomy, University of St Andrews, North Haugh, St Andrews, Fife, KY16 9SS, UK;\\
$^{7}$ Astrophysics Research Institute, Liverpool John Moores University, IC2, Liverpool Science Park, 146 Brownlow Hill, Liverpool L3 5RF, UK;\\
$^{8}$ Centre for Astrophysics \& Supercomputing, Swinburne University of Technology, Mail H29, PO Box 218, Hawthorn, VIC 3122, Australia\\
$^{9}$ Research School of Astronomy and Astrophysics, The Australian National University, Canberra, ACT 2611, Australia\\
$^{10}$ European Southern Observatory, Karl-Schwarzschild-Str. 2, 85748 Garching bei M\"unchen, Germany\\
$^{11}$ School of Physics, The University of Melbourne, VIC 3010, Australia\\
$^{12}$ Institute for Astronomy, University of Hawaii, 2680 Woodlawn Drive, Honolulu, HI 96822, USA\\
$^{13}$ Institute of Photonics and Optical Science (IPOS), School of Physics, The University of Sydney, NSW 2006, Australia\\
$^{14}$ Leibniz-Institut f\"ur Astrophysik Potsdam (AIP), An der Sternwarte 16, D-14482 Potsdam, Germany \\
$^{15}$ Department of Physics \& Astronomy, Rutgers University, Piscataway, NJ 08854, USA\\
$^{16}$ Department of Physics, University of Oxford, Denys Wilkinson Building, Keble Rd., Oxford, OX1 3RH, UK\\
$^{17}$ Institute for Computational Cosmology, Department of Physics, Durham University, South Road, Durham DH1 3 LE, U.K.\\
$^{18}$ Department of Physics \& Astronomy, University of North Carolina, Chapel Hill, NC 27559, USA\\
$^{19}$ Indian Institute of Science Education and Research Mohali-IISERM, Knowledge City, Sector 81, Manauli, PO 140306, India\\
$^{20}$ Department of Physics and Astronomy, Macquarie University, NSW 2109,
Australia\\
$^{21}$ Physics Department, Durham University, South Road, Durham, DH1 3LE, England.\\
$^{22}$ Visiting Professor, Sydney Institute for Astronomy (SIfA), School of Physics, The University of Sydney, NSW 2006, Australia \\


\begin{thebibliography}{}
\bibitem[Abazajian et al.(2009)]{Aba09} Abazajian K. N., et al., 2009, APJS, 182, 543
\bibitem[Allen et al.(2014)]{All14} Allen J. et al., MNRAS, 446, 1567.
\bibitem[Alpaslan et al.(2012)]{Alp12} Alpaslan M., et al., 2012, MNRAS, 426, 2832
\bibitem[Alpaslan et al.(2014a)]{Alp14a} Alpaslan M., et al., 2014a, MNRAS, 438, 177
\bibitem[Alpaslan et al.(2014b)]{Alp14b} Alpaslan M., et al., 2014b, MNRAS, 440, 106
\bibitem[Baldry et al.(2008)]{Bal08} Baldry I. K., et al., 2008, MNRAS, 388, 945
\bibitem[Baldry et al.(2010)]{Bal10} Baldry I. K., et al., 2010, MNRAS, 404, 86
\bibitem[Baldry et al.(2012)]{Bal12} Baldry I. K., et al., 2012, MNRAS, 421, 621
\bibitem[Becker, White \& Helfand(1995)]{Bec95} Becker R. H., White R. L., \& Helfand D. J. 1995, ApJ, 450, 559
\bibitem[Bekki(2009)]{Bek09} Bekki K., 2009, MNRAS, 399, 2221
\bibitem[Bekki(2014)]{Bek14} Bekki K., 2014, MNRAS, 438, 444
\bibitem[Best(2004)]{Bes04} Best P. N., 2004, MNRAS, 351, 70
\bibitem[Blanc et al.(2013)]{Bla13} Blanc G. A., et al., 2013, AJ, 145, 138 
\bibitem[Bland-Hawthorn et al.(2011)]{JBH2011} Bland-Hawthorn J. et al., 2011, Optics Express, 19, 2649
\bibitem[Bravo-Alfaro, van Gorkom, \& Caretta(2008)]{bravo2008} Bravo-Alfaro H., van Gorkom J.~H., Caretta C., 2008, in Problems of Practical Cosmology, Volume 1, Baryshev Y.~V., Taganov I.~N., Teerikorpi P., eds., pp. 102--105
\bibitem[Brooks et al.(2009)]{Bro09} Brooks A.~M., Governato F., Quinn T., Brook C.~B., Wadsley J., 2009, ApJ, 694, 396

\bibitem[Brough et al.(2013)]{Bro13} Brough S., et al., 2013, MNRAS, 435, 2903
\bibitem[Brough et al.(2014)]{Bro14} Brough S., et al., 2014, AAO Observer, in press.
\bibitem[Bruzual(1993)]{Bru93} Bruzual G., 1993, ApJ, 273, 105
\bibitem[Bruzual \& Charlot(2003)]{Bru03} Bruzual G., Charlot S., 2003, MNRAS, 344, 1000
\bibitem[Bryant et al.(2011)]{JB2011} Bryant J. J., O'Byrne J. W., Bland-Hawthorn J., Leon- Saval S. G., 2011, MNRAS, 797
\bibitem[Bryant et al.(2012)]{JB2012} Bryant J. J., et al., 2012, McLean I. S., Ramsay S. K., Takami H., eds., in Ground-based and Airborne Instrumentation for Astronomy IV. Proceedings of the SPIE, Volume 8446, article id. 84460X.
\bibitem[Bryant et al.(2014)]{JB2014} Bryant J. J., Bland-Hawthorn J., Fogarty L. M. R., Lawrence J. S., Croom S. M., 2014, MNRAS, 438, 869
\bibitem[Cappellari \& Copin(2003)]{Cap03} Cappellari M., Copin Y., 2004, MNRAS, 342, 345
\bibitem[Cappellari et al.(2011)]{Cap11} Cappellari M. et al., 2011a, MNRAS, 413, 813
\bibitem[Cappellari et al.(2012)]{Cap2012} Cappellari M., et al., 2012, Nature, 484, 485
\bibitem[Cappellari et al.(2013)]{Cap13} Cappellari M. et al., 2013 ApJ, 778, L2
\bibitem[Carlberg et al.(1997)]{Car97} Carlberg, R. G et al., 1997, ApJ, 485, 13
\bibitem[Catinella et~al.(2013)]{catinella13} Catinella B., et al., 2013, MNRAS, 436, 34
\bibitem[Cluver et al.(2014)]{Clu14} Cluver M. E., et al, 2014, ApJ, 782, 90
\bibitem[Colless et~al.(2001)]{Col01} Colless M. et al., 2001, MNRAS, 328, 1039
\bibitem[Condon et al.(1998)]{Con98} Condon J. J., Cotton W. D., Greisen E. W., Yin Q. F., Perley R. A., Taylor G. B., Broderick J. J., 1998, AJ, 115, 1693
\bibitem[Cortese et al.(2011)]{cortese11} Cortese L., Catinella B., Boissier S., Boselli A., Heinis S., 2011, MNRAS, 415, 1797
\bibitem[Croom et al.(2012)]{Cro2012} Croom S., et al., 2012, MNRAS 421, 872 
\bibitem[da Cunha, Charlot \& Elbaz(2008)]{daC08} da Cunha E., Charlot S., Elbaz D., 2008, MNRAS, 388, 1595
\bibitem[de Zeeuw et al.(2002)]{deZ02} de Zeeuw P. T., et al., 2002, MNRAS, 329, 513
\bibitem[de Jong et al.(2013)]{deJ13} de Jong J.T.A., et al., 2013, Experimental Astronomy, 35, 25
\bibitem[di Cinto et al.(2014)]{diC14} di Cintio A.,  Brook C. B., Macci\`{o} A. V., Stinson G. S., Knebe A., Dutton A. A., Wadsley J., 2014, MNRAS, 437, 415
\bibitem[Dressler(1980)]{Dre80} Dressler A., 1980, ApJ, 236, 351
\bibitem[Drinkwater et al.(2010)]{Dri10} Drinkwater M. J. et al., 2010, MNRAS, 401, 1429
\bibitem[Driver et al.(2009)]{Dri09} Driver S.P., Norberg P., Baldry I. K., Bamford S. P., Hopkins A. M., Liske J., Loveday J., Peacock J. A., 2009, A\&G, 50, 5.12
\bibitem[Driver et al.(2011)]{Dri11} Driver S. P. et al., 2011, MNRAS, 413, 971
\bibitem[Dutton et al.(2011)]{Dut11} Dutton A. A. et al., 2011, MNRAS, 416, 322
\bibitem[Eales et al.(2010)]{Eal10} Eales, S. et al., 2010, PASP, 122, 499
\bibitem[Eke et al.(2005)]{Eke05} Eke V. et al., 2005, MNRAS, 362, 1233
\bibitem[Emsellem et al.(2011)]{Ems11} Emsellem E., et al. 2011, MNRAS, 414, 888
\bibitem[Fogarty et al.(2012)]{Fog2012} Fogarty L. M. R. et al., 2012 ApJ, 761, 169
\bibitem[Fogarty et al.(2014)]{Fog2014} Fogarty L. M. R. et al., 2014 MNRAS, 443, 485
\bibitem[Garc\'{i}a-Benito et al.(2014)]{Gar14} Garc\'{i}a-Benito R. et al., 2014, arXiv:1409.8302 
\bibitem[Giovanelli et al.(2005){Giovanelli}]{alfaalfa05} Giovanelli R., et al., 2005, AJ, 130, 2598
\bibitem[Gnedin \& Kravtsov(2010)]{Gne10} Gnedin N. Y., Kravtsov A. V., 2010, ApJ, 714, 287
\bibitem[Governato et al.(2012)]{Gov12} Governato F. et al., 2012, MNRAS, 422, 1231
\bibitem[Hambly et al.(2001)]{Ham2001} Hambly N. C., et al., 2001, MNRAS, 326, 1279
\bibitem[Han et al.(2014)]{Han14} Han J., et al., 2014, arXiv:1404.6828
\bibitem[Haynes et~al.(2011)]{haynes2011} Haynes M.~P., et al., 2011, AJ, 142, 170
\bibitem[Heiderman et al.(2011)]{Hei11} Heiderman A. L., Evans N. J. II, Gebhardt K., Blanc G., Davis T. A., Papovich C., Iono D., Yun M. S., 2011, Proc. of the Frank N. Bash Symposium, 2011, 29, Ed. S. Salviander, J. Green, and A. Pawlik
\bibitem[Hill et al.(2011)]{Hil11} Hill D., et al., 2011, MNRAS, 412, 765
\bibitem[Hinshaw et al.(2009)]{Hin09}Hinshaw G. et al.,  2009, ApJS, 180, 225
\bibitem[Ho et al.(2014)]{Ho14} Ho I. et al., 2014, 444, 3894
\bibitem[Hopkins, Quataert, \& Murray(2012)]{Hop12} Hopkins P. F., Quataert E., Murray N., 2012, MNRAS, 421, 3522
\bibitem[Hopkins et al.(2013)]{Hop13} Hopkins A., et al., 2013, MNRAS, 430, 2047
\bibitem[Iono, Yun \& Mihos(2004)]{Ion04} Iono D., Yun M. S., Mihos J. C., 2004, ApJ 616, 199
\bibitem[Jarrett et al.(2012)]{Jar12} Jarrett T., et al., 2012, AJ, 144, 68
\bibitem[Johnston et al.(2007)]{Joh07} Johnston S., et al., 2007, Publications of the Astronomical Society of Australia, 24, 174
\bibitem[Krajnovi\'{c} et al.(2011)]{Kra11} Krajnovi\'{c} D., et al., 2011, MNRAS 414, 2923
\bibitem[Kelvin et al.(2012)]{Kel12} Kelvin L.S., et al., 2012, MNRAS, 421, 1007
\bibitem[Kere{\v s} et al.(2005)]{Ker05} Kere{\v s} D., Katz N., Weinberg D.~H., 
Dav{\'e} R., 2005, MNRAS, 363, 2
\bibitem[Koopmann \& Kenney(2004)]{Koo04} Koopmann R. A., Kenney J. D. P., 2004, ApJ, 613, 866 
\bibitem[Kron(1980)]{Kro80} Kron R. G., 1980, ApJS, 43, 305
\bibitem[Krumholz \& Dekel(2012)]{Kru12} Krumholz M. R., Dekel A., 2012, ApJ, 753, 16
\bibitem[Lawrence et al.(2007)]{Law07} Lawrence A. et al., 2007, MNRAS, 379, 1599

\bibitem[Le F\`{e}vre et al.(2004)]{LeF04} Le Fevre O., et al., 2004, A\&A, 428, 1043
\bibitem[Lewis et al.(2002)]{Lew02} Lewis I. J., et al., 2002, MNRAS, 334, 673
\bibitem[Liske et al.(2003)]{Lis03} Liske J., Lemon D. J., Driver S. P., Cross N. J. G., Couch W. J., 2003, MNRAS, 344, 307
\bibitem[Lorente(2014)]{Lor2014} Lorente N. P. F., 2014, in Manset N. and Forshay P., eds., in ASP conference series, vol. 485,  Astronomical Data Analysis Software and Systems XXIII, Astron. Soc. Pac., San Francisco, p. 95 
\bibitem[Martin et al.(2005)]{Mar05a} Martin C.D., et al., 2005, ApJ, 619, 1
\bibitem[Martin(2005b)]{Mar05b} Martin C. L., 2005, ApJ, 621, 227
\bibitem[Mateus \& S\'{o}dre(2003)]{Mat03} Mateus Jr. A., Sodr«e Jr. L., 2004, MNRAS, 349, 1251
\bibitem[Mauch et al.(2013)]{Mau13} Mauch T., et al., 2013, 435, 650
\bibitem[Mauch et al.(2003)]{Mau03} Mauch T., Murphy T., Buttery H. J., Curran J., Hunstead R.W., Piestrzynski B., Robertson J. G., Sadler E. M., 2003, MNRAS, 342, 1117
\bibitem[Moore, Lake \& Katz(1998)]{Moo98} Moore, B., Lake, G., \& Katz, N. 1998, ApJ, 495, 139
\bibitem[Ortiz et al.(2011)]{Ort11} Ortiz I., et al., 2011, arXiv:1110.0503
\bibitem[Owers et al.(2009)]{Owe09} Owers M. S., Nulsen P. E. J., Couch W. J., Markevitch M., Poole G. B., 2009, ApJ, 692, 702
\bibitem[Robertson \& Kravtsov(2008)]{Rob08} Robertson B. E., Kravtsov A. V., 2008, ApJ, 680, 1083
\bibitem[Robotham et al.(2010)]{Rob10} Robotham A.~S.~G., et al., 2010, PASA, 27, 76
\bibitem[Robotham et al.(2011)]{robotham11} Robotham A.~S.~G., et al., 2011, MNRAS, 416, 2640
\bibitem[Rosales-Ortega et al.(2010)]{Ros10} Rosales-Ortega F. F., Kennicutt R. C., S‡nchez S. F., D'az A. I., Pasquali A., Johnson B. D., Hao C. N., 2010, MNRAS, 405, 735
\bibitem[Roth et al.(2005)]{Rot05} Roth M. M. et al., 2005, PASP, 117, 620
\bibitem[Saintonge(2007)]{saintonge07} Saintonge A., 2007, AJ, 133, 2087
\bibitem[S\'{a}nchez et al.(2012)]{San12} S\'{a}nchez S. F., et al., 2012, A\&A, 538, A8
\bibitem[Scoville et al.(2007)]{Sco07} Scoville N., et al., 2007, ApJS, 172, 1
\bibitem[Shanks et al.(2013)]{Sha13} Shanks T. et al., 2013, ESO Messenger, 154, 38
\bibitem[Sharp et al.(2014)]{Sha14} Sharp R., et al. 2015, MNRAS, 446, 1551
\bibitem[Sharp et al.(2006)]{Sha06} Sharp R., et al., 2006, in McLean I.S., Iye W., eds, SPIE Conf. Ser. Vol. 6269, id. 62690G, in Ground based and Airbourne Instrumentation for Astronomy
\bibitem[Taylor et al.(2011)]{Tay11} Taylor E. N. et al., 2011, MNRAS, 418, 1587
\bibitem[Tonry et al.(2000)]{Ton00} Tonry J. L., Blakeslee J. P., Ajhar E. A., Dressler A., 2000, ApJ,
530, 625
\bibitem[Tremaine et al.(2002)]{Tre02} Tremaine S. et al., 2002, ApJ, 574, 740
\bibitem[Tully \& Fisher(1977)]{Tull77} Tully R. B., Fisher J. R., 1977, A\&A, 54, 66
\bibitem[Volonteri, Haardt, \& Madau(2003)]{Vol03} Volonteri M., Haardt F., Madau P., 2003, ApJ, 582, 559
\bibitem[Walcher et al.(2014)]{Wal14} Walcher C. J. et al., 2014, A\&A, 569, 1
\bibitem[Wijesinghe et al.(2012)]{Wij12}  Wijesinghe D. B. et al.,  2012, MNRAS, 423, 3679
\bibitem[York et al.(2000)]{Yor2000} York D. G. et al., 2000, AJ, 120, 1579
\bibitem[Zacharias et al.(2013)]{Zac13} Zacharias N., Finch C. T., Girard T. M., Henden A., Bartlett J. L., Monet D. G., Zacharias M. I., 2013, AJ, 145, 44
\bibitem[Zavala et al.(2009)]{Zav09} Zavala J., Jing Y. P., Faltenbacher A., Yepes G., Hoffman Y., Gottl\"{o}ber S., Catinella B., 2009, ApJ, 700, 1779

\end{thebibliography}
\end{document}